\documentclass{article}

\usepackage{arxiv}

\usepackage[utf8]{inputenc} % allow utf-8 input
\usepackage[T1]{fontenc}    % use 8-bit T1 fonts
\usepackage{hyperref}       % hyperlinks
\usepackage{url}            % simple URL typesetting
\usepackage{booktabs}       % professional-quality tables
\usepackage{amsfonts}       % blackboard math symbols
\usepackage{nicefrac}       % compact symbols for 1/2, etc. 
\usepackage{microtype}      % microtypography
\usepackage{amsmath}
\usepackage{cleveref}       % smart cross-referencing
\usepackage{lipsum}         % Can be removed after putting your text content
\usepackage{graphicx}
\usepackage{natbib}
\usepackage{doi}
\usepackage{amssymb}
\usepackage{caption}
\usepackage{subcaption}
\DeclareMathOperator{\sech}{sech}

\title{Breather bound states in a parametrically driven magnetic wire}

\newif\ifuniqueAffiliation
\ifuniqueAffiliation % Standard variant of author block 

\author{ \href{https://orcid.org/0000-0000-0000-0000}{\includegraphics[scale=0.06]{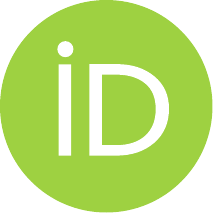}\hspace{1mm}Camilo José Castro}\thanks{Use footnote for providing further
		information about author (webpage, alternative
		address)---\emph{not} for acknowledging funding agencies.} \\
	Instituto de Alta Investigación\\ 
	Universidad de Tarapacá \\
	Casilla 7D, Arica 1000000, Chile\\
	Carrera de Física\\
	Universidad Mayor de San Andrés \\
	La Paz, Bolivia \\
	\texttt{hippo@cs.cranberry-lemon.edu} \\
	\And
	\href{https://orcid.org/0000-0000-0000-0000}{\includegraphics[scale=0.06]{orcid.pdf}\hspace{1mm}Elias D.~Striatum} \\
	Department of Electrical Engineering\\
	Mount-Sheikh University\\
	Santa Narimana, Levand \\
	\texttt{stariate@ee.mount-sheikh.edu} \\
}
\else
\usepackage{authblk}

\newbox{\orcid}\sbox{\orcid}{\includegraphics[scale=0.06]{orcid.pdf}} 
\author[1,2]{%
	\href{https://orcid.org/0000-0003-0968-5715}{\usebox{\orcid}\hspace{1mm}Camilo José Castro\thanks{\texttt{ccastro@fiumsa.edu.bo}}}%
}
\author[3]{%
	\hspace{1mm}Ignacio Ortega-Piwonka%
} 

\author[4]{%
	\hspace{1mm}Boris A. Malomed%
} 

\author[2]{%
	\hspace{1mm}Deterlino Urzagasti%
} 

\author[5]{%
	\hspace{1mm}Liliana Pedraja-Rejas%
} 

\author[6]{%
	\hspace{1mm}Pablo Díaz%
} 

\author[1]{%
	\hspace{1mm}David Laroze\thanks{\texttt{dlarozen@academicos.uta.cl}}%
} 

\affil[1]{Instituto de Alta Investigación, Universidad de Tarapacá 
	Casilla 7D, Arica 1000000, Chile}
\affil[2]{Carrera de Física, Universidad Mayor de San Andrés, La Paz, Bolivia } 
\affil[3]{Grupo de Dinámica No Lineal, Caos y Sistemas Complejos, Universidad Rey Juan Carlos, Tulipán s/n, Móstoles 28933, Spain } 
\affil[4]{Department of Physical Electronics, School of Electrical Engineering, Faculty of Engineering, and Center for Light-Matter Interaction, Tel Aviv University, Tel Aviv 69978, Israel} 
\affil[5]{Departamento de Ingeniería Industrial y de Sistemas, Universidad de Tarapacá, Casilla 7D, Arica 1000000, Chile} 
\affil[6]{Departamento de Ciencias Físicas, Universidad de La Frontera, Casilla 54-D, Temuco, Chile} 

\fi

\renewcommand{\shorttitle}{Breather bound states in a parametrically driven magnetic wire}

\hypersetup{
pdftitle={A template for the arxiv style},
pdfsubject={q-bio.NC, q-bio.QM},
pdfauthor={David S.~Hippocampus, Elias D.~Striatum},
pdfkeywords={First keyword, Second keyword, More},
}

\begin{document}
\maketitle

\begin{abstract}
	We report the results of systematic investigation of localized dynamical states in the model of a one-dimensional magnetic wire, which is based on the Landau-Lifshitz-Gilbert (LLG) equation. The dissipative term in the LLG equation is compensated by the parametric drive imposed by the external AC magnetic field, which is uniformly applied perpendicular to the rectilinear wire. The existence and stability of the localized states is studied in the plane of the relevant control parameters, \textit{viz}., the amplitude of the driving term and the detuning of its frequency from the parametric resonance. With the help of systematically performed simulations of the LLG equation, existence and stability areas are identified in the parameter plane for several species of the localized states: stationary single- and two-soliton modes, single and double breathers, drifting double breathers with spontaneously broken inner symmetry, and multi-soliton complexes. Multistability occurs in this system. The breathers emit radiation waves (which explains their drift caused by the spontaneous symmetry breaking, as it breaks the balance between the recoil from the waves emitted to left and right), while the multi-soliton complexes exhibit cycles of periodic transitions between three-, five-, and seven-soliton configurations. Dynamical characteristics of the localized states are systematically calculated too. These include, in particular, the average velocity of the
asymmetric drifting modes, and the largest Lyapunov exponent, whose negative and positive values imply that the intrinsic dynamics of the respective modes is regular or chaotic, respectively.
\end{abstract}

% keywords can be removed
\keywords{Landau-Lifshitz equation \and dispersive radiation \and soliton dynamics; Lyapunov exponents\and multistability}

\section{Introduction}
Pattern formation in nonlinear dissipative media relies on the coexistence
of two balance conditions: between the nonlinear self-focusing and linear
self-stretching, under the action of diffraction and/or dispersion, and
between dissipative losses and a compensating mechanism. In physical systems
governed by complex Ginzburg-Landau equations, the losses are balanced by
the intrinsic gain \cite{Aranson-Kramer,Rosanov}, while in passively driven
optical cavities the compensation is provided by an external pump, as
modeled by the Lugiato-Lefever equation \cite{LL,LL2}.

The parametrically-driven damped nonlinear Schr\"{o}dinger (PDNLS) equation
\cite{miles1984,Barash-1} is known as a universal model of periodically
forced dissipative systems. Solutions of this equation are known to produce
various dynamical regimes, including stationary, periodic, and chaotic ones,
such as Faraday waves \cite%
{faraday1831xvii,scott1969nonlinear,coullet1994dispersion,
clerc2010interaction}, single solitons \cite{Barash-1,barashenkov2011time,alexeeva2000impurity,
barashenkov1999stable,zemlyanaya2009oscillating}, two-soliton states \cite%
{barashenkov2011soliton,Urzagasti_et_al_JAppPhys_2012}, as well as
spatiotemporal chaos \cite{Barash-1,shchesnovich2002soliton}. The parametric
instability of a vertically vibrated Newtonian fluid, modeled by the PDNLS
equation, leads to notable hydrodynamic phenomena manifested by standing
(Faraday) waves on its surface \cite{faraday1831xvii}. These standing waves
are most sensitive to the parametric forcing at half its frequency (the $2:1$
parametric resonance) \cite{arnold2012geometrical}. In weakly dissipative
systems, the PDNLS equation captures the development of the parametric
instability occurring near the $2:1$ resonance \cite%
{Clerc_Coulibali_Laroze_IntJBifChaos_19_3525}. In this context, it is
relevant to mention that stabilization of dissipative solitons by the
parametric drive was studied in various contexts, such as the periodically
time-modulated damped nonlinear Schr\"{o}dinger equation for wave amplitudes
\cite{Okamura_Konno_Japan_1989,Barash-1}, and chains of coupled pendula \cite%
{pendula}, just to mention a few. Chaotic dynamical
states generated by the PDNLS equation have been studied in detail too \cite%
{Barash-2,Barash-3}. Also investigated were bound states of solitons in the
same model, chimeras and localized dynamical chaos in
parametrically-driven nonlinear lattices in the discrete version of the
equation \cite{chimeras}, as well as periodic waves and multistable
dissipative solitons produced by the parametrically-driven version of the
complex Ginzburg-Landau equation (in other words, the PDNLS equation with
complex coefficients in front of the dispersion and nonlinearity terms) \cite%
{CGL,HS}. Patterns and localized structures produced by (generalized) PDNLS
and related equations can be found in Refs. \cite
{reyes2024characterization,leon2015traveling,leon2024faraday,moille2024parametrically, bogdan2022structure,shaukat2022spatial,cabanas2021quasi,urzagasti2017two,urzagasti2014localized,
marin2023drifting,barbosa2023artificial,dileep2023emergent,parra2022dissipative, englebert2021parametrically,diamantidis2021exciting,yamaguchi2021generation, 
mertens2020empirical,barashenkov2020stable,urra2019localized,edri2020spatial, 
ferre2017localized,clerc2014propagative}.

Among various physical settings, nonlinear phenomena are common in magnetic
materials. In the classical regime, the appropriate model for the
magnetization dynamics is provided by the Landau-Lifshitz-Gilbert (LLG)
equation \cite{NL_Magnetization} and its generalizations \cite{LLG+ac,
Lakshmanan}. In the case of magnetic nanoparticles, where magnetization is
represented by a single magnetic domain, both theoretical and experimental
studies have explored different routes to periodicity, chaos, and
multistability \cite%
{montoya2019magnetization,bragard2021study,alvarez2000quasiperiodicity,
perez2015effect,smith2010period,leon2024faraday,velez2020periodicity,ferona2017nonlinear, smith2009nonlinear,sementsov2009chaotic,botha2023chaotic,shen2024skyrmion,yamaguchi2023computational}%
. A relevant possibility is to include additional terms in the equation that
account for magnetic inertia \cite%
{unikandanunni2022inertial,rodriguez2024spin}. In the case of objects where
spatial dimensions play a fundamental role, such as wires (both straight and
curved ones), tubes, membranes, cubes, tori, etc., a wide variety of
magnetic textures and a broad range of pattern-formation regimes have been
discovered. Among them, well-known examples are vortices, skyrmions, and
droplets \cite%
{jain2012chaos,pivano2016chaotic,guslienko2010nonlinear,ovcharov2024antiferromagnetic, d2023micromagnetic,gareeva2023nutation,fert2017magnetic,jiang2024magnetic,garcia2016skyrmion,
jiang2017skyrmions,deng2022observation,kosevich1990magnetic,mohseni2013spin,roessli2001formation,
rothos2023dissipative}.

Two classes of physically relevant states in dissipative magnetic media,
where losses can be counterbalanced by the parametric drive in the form of a
periodically time-modulated (AC) external magnetic field, are magnetic
solitons and Faraday waves. In this scenario, the parametrically-driven LLG
equation near the $2:1$ parametric resonance can be approximated by the
PDNLS equation \cite{Barash-1}. Therefore, all the phenomena predicted by
the PDNLS equation can be also observed in the framework of its LLG
counterpart. In particular, standing waves, pulses, solitons, breathers, and
the interaction of in-phase and out-of-phase stationary solitons in magnetic
wires, as well as solitons and magnetic textures in the magnetic plane, have
been studied in this parametrically driven context \cite%
{clerc2010interaction,Clerc_Coulibali_Laroze_PhysRevE_2008,
Clerc_Coulibali_Laroze_IntJBifChaos_19_2717,Clerc_Coulibali_Laroze_IntJBifChaos_19_3525,Clerc_Coulibali_Laroze_PhysicaD_2010,Urzagasti2013}. However, the previous studies of PDNLS-based models did not address the
complexity of non-stationary \emph{bound states of solitons}, which is an
essential direction for the extension of the work on this topic.

This work aims to tackle this possibility, addressing novel dynamical
scenarios involving the interaction of two one-dimensional solitons in a
parametrically driven magnetic chain, treated in the continuum
approximation. Starting from the parametrically-driven LLG model, we
demonstrate that the interacting solitons either merge into a single mode or
form a clearly identified bound state. Previously unexplored complex
dynamical states emerge as an extension of the basic soliton dynamics.
Specifically, by varying the model parameters, we observe that standing
waves exhibit an oscillatory instability. Further changes in the parameters
lead to a secondary oscillatory instability, which gives rise to spontaneous
symmetry breaking of double breathers, which is accompanied by their drift,
and, eventually, spatiotemporal chaos. The dynamics are quantified by dint
of several indicators, including bifurcation diagrams and Lyapunov exponents.

The manuscript is structured as follows: Sections \ref%
{section_Theoretical_model} and \ref{section_Numerical_framework} present,
respectively, the model and scheme of the adopted numerical analysis.
Dynamical indicators, used for the characterization of various novel states
produced by the model, are introduced in Section \ref%
{section_Dynamical_indicators}. Systematic numerical results and the
corresponding discussion are reported in Section \ref{section_Results}, and
the paper is concluded by Section \ref{section_Conclusions}.

\section{The theoretical model}

\label{section_Theoretical_model}

\begin{figure}[tbp]
\centering
\begin{subfigure}{0.485\textwidth}
        \includegraphics[width=\linewidth]{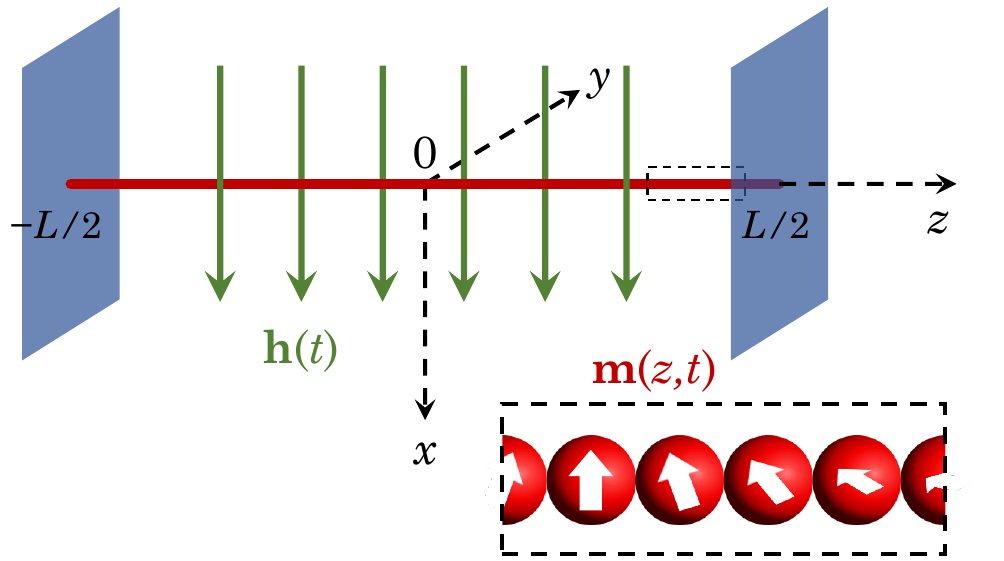}
        \caption{\centering}
    \end{subfigure}
\begin{subfigure}{0.485\textwidth}
        \includegraphics[width=\linewidth]{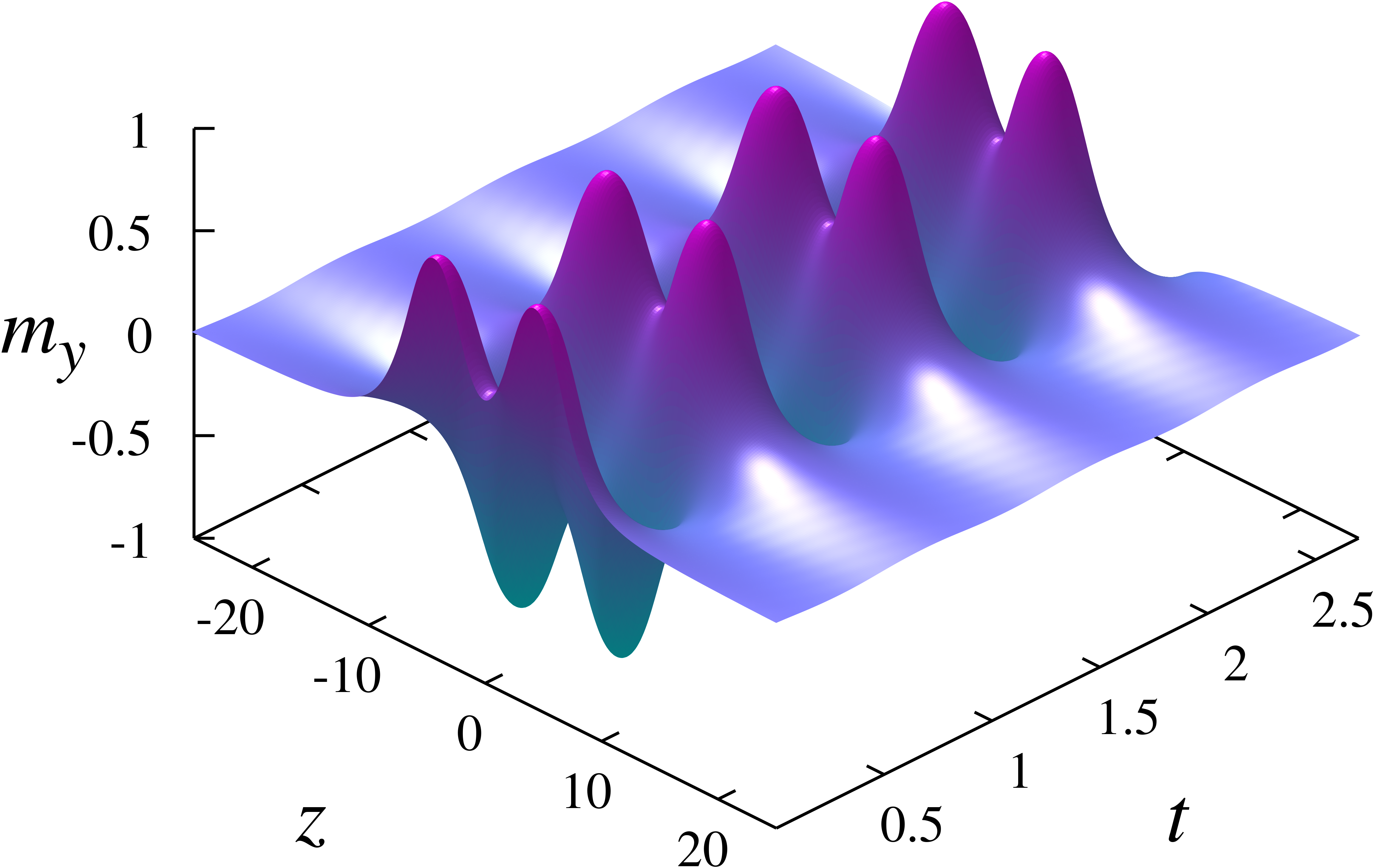}
        \caption{\centering}
    \end{subfigure}
\vskip\baselineskip
\begin{subfigure}{0.485\textwidth}
        \includegraphics[width=\linewidth]{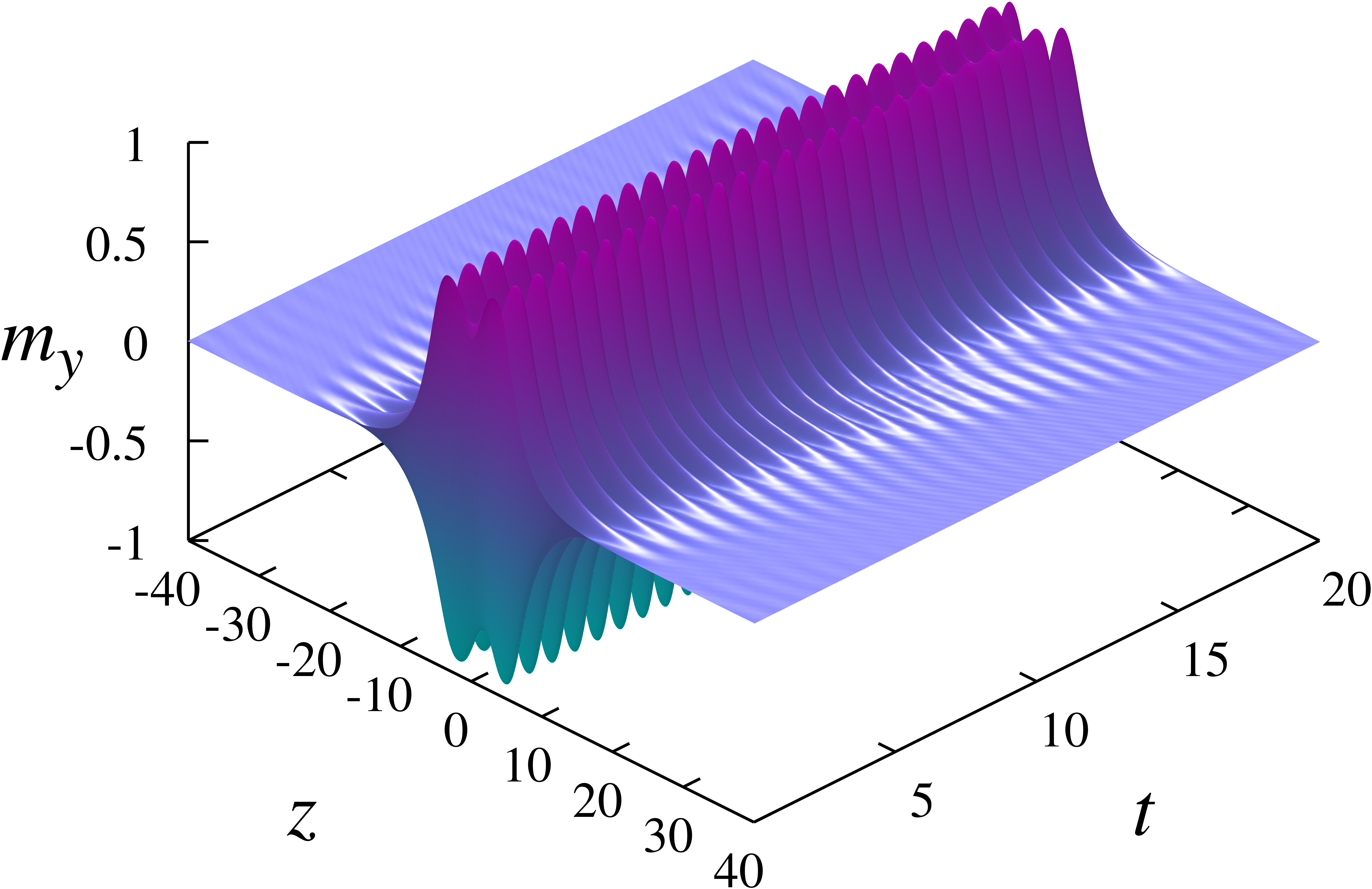}
        \caption{\centering}
    \end{subfigure}
\begin{subfigure}{0.485\textwidth}
        \includegraphics[width=\linewidth]{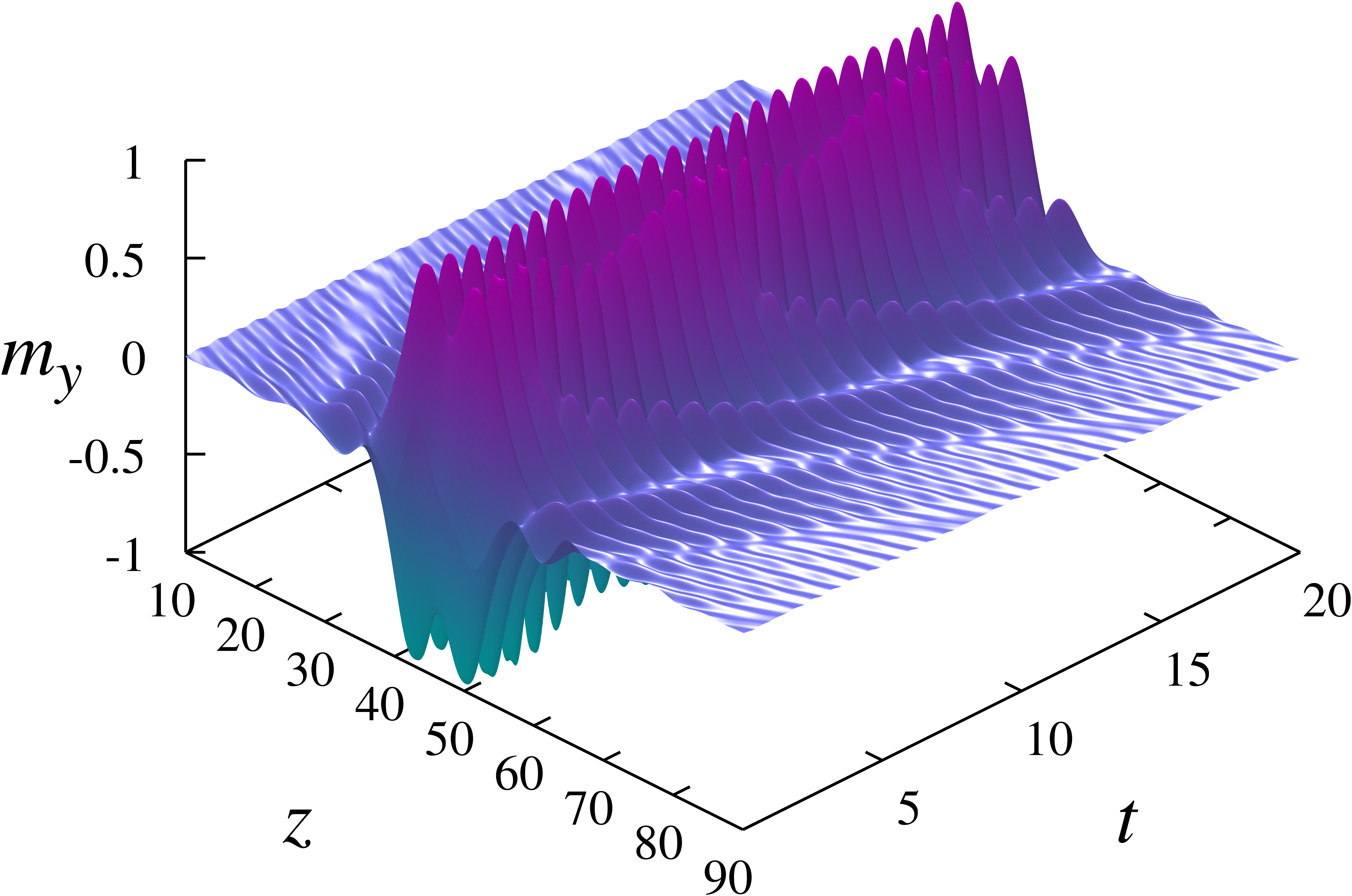}
        \caption{\centering}
    \end{subfigure}
\caption{(color online) (\textbf{a}) The schematic of the one-dimensional
wire aligned with the $z$-axis. The wire is made of magnetic particles
subjected to the action of the perpendicular time-modulated magnetic field, $%
\mathbf{h}(t)$. As a result, the magnetization of the constituent particles
(schematically denoted by white arrows in the inset), which is represented
by continuous field $\mathbf{m}(z,t)$, is polarized according to LLG
equation (\protect\ref{eq_LLG}). (\textbf{b--d}) Simulations of Eq. (\protect
\ref{eq_LLG}) showing the double-soliton response for different values of
the modulation parameters, namely: (\textbf{b}) $(\protect\nu %
,h_{0})=(-0.3,0.55)$ (the standard double soliton); (\textbf{c}) $(\protect%
\nu ,h_{0})=(-0.44,0.65)$ (a symmetric breathing double soliton); (\textbf{d}%
) $(\protect\nu ,h_{0})=(-0.402,0.636)$ (an asymmetric breathing double
soliton). }
\label{fig_double_soliton}
\end{figure}

We consider a magnetic wire, with the normalized magnetization field $%
\mathbf{m}=\mathbf{m}(\mathbf{r},t)$, where $\mathbf{r}$ and $t$ represent
the spatial coordinates and time, respectively. We concentrate on a
ferromagnetic anisotropic long wire, assuming that the propagation of
magnetization waves along the wire axis, i.e., in the direction represented
by $\mathbf{\hat{z}}=(0,0,1)$ \cite{Aharoni_Ferromagnetism}, is governed by
the LLG equation. In particular, this equation was used to model a
one-dimensional magnetic wire driven by a time-modulated magnetic field
applied in the perpendicular direction, to study the transition from regular
solitons to regular bound states \cite{Urzagasti_et_al_JAppPhys_2012}. Here
we employ the same setup, as shown in Fig. \ref{fig_double_soliton}(a). For
this setting, the LLG equation can be cast in the form of
\begin{equation}
\frac{\partial \mathbf{m}}{\partial t}=-\mathbf{m}\times \mathbf{\Gamma }%
+\alpha \mathbf{m}\times \frac{\partial \mathbf{m}}{\partial t}.
\label{eq_LLG}
\end{equation}%
where $\mathbf{\Gamma }$ is the effective torque and $\alpha $ is the
damping coefficient. This equation is derived from the general LLG model by
taking into regard that the scaled magnetization vector, $\mathbf{m}$, which
represents the local orientation of constituent particles of the medium,
keeps a constant absolute value, $|\mathbf{m}|^{2}=1$ and assuming the
one-dimensional evolution of the $\mathbf{m}$ field along coordinate $z$. In
this case, the effective torque in Eq. (\ref{eq_LLG}), acting upon the
magnetization in a long wire, may be approximated as
\begin{equation}
\Gamma =\frac{\partial ^{2}\mathbf{m}}{\partial z^{2}}-\beta (\mathbf{m}%
\cdot \hat{\mathbf{z}})\hat{\mathbf{z}}+\mathbf{h}.  \label{eq_Gamma}
\end{equation}%
It includes scaled parameter $\beta $ accounting for the anisotropy along
the $z$-axis, which depends on properties of the magnetic medium, and the
external modulated magnetic field composed of constant (DC) and
time-modulated (AC) terms, \textit{viz}.,
\begin{equation}
\mathbf{h}(t)=(h_{c}+h_{0}\cos (\Omega t))\hat{\mathbf{x}},  \label{eq_h}
\end{equation}%
where $h_{c}$, $h_{0}$, $\Omega $, and $\hat{\mathbf{x}}$ are the modulation
offset, AC amplitude. frequency, and the unit vector perpendicular to the
wire, respectively \cite{Urzagasti_et_al_JAppPhys_2012}.

In the application to a single magnetic particle, which does not admit
spatial dependence, the LLG equation exhibits a variety of stable
equilibrium solutions, including the fixed attractor with $\mathbf{m}=\hat{%
\mathbf{x}}$, when the modulation is null (i.e., $h_{0}=0$). This means that
the magnetization is aligned to the DC magnetic field. When the AC field is
present, one can find periodic, quasi-periodic, and chaotic responses \cite%
{Laroze_IEEE_2011, Laroze_IEEE_2012}. In particular, small perturbations
around the stationary attractor produce damped oscillations with frequency $%
\Omega _{0}=\sqrt{h_{c}(h_{c}+\beta )}$. The attractor becomes unstable via
an oscillatory instability when the magnetic field is modulated with a
frequency close to the parametric-resonance value,
\begin{equation}
\Omega =2(\Omega _{0}+\nu ),  \label{eq_detuning}
\end{equation}%
where $\nu $ is the detuning from the resonance. Specifically, this
bifurcation occurs when the modulation amplitude surpasses a critical value
related to $\nu $ by the following equation:

\begin{equation}  
(h_0^{\text{crit}})^2 = 
\bigg(\frac{4\Omega_0}{\beta}\bigg)^2 \bigg(\nu^2 +
\Big(\frac{\alpha(\beta+2h_c)}{2}\Big)^2\bigg).  \label{eq_Arnold}
\end{equation}

Under this condition, the magnetic particle responds by subharmonic
oscillations along the $x$ and $y$-axis, directly exhibiting the parametric
resonance \cite{Clerc_Coulibali_Laroze_PhysicaD_2010}. Equation (\ref%
{eq_Arnold}) corresponds to a hyperbola in the $(\nu ,h_{0})$ plane, which
represents the first Arnold tongue in the system.

\unskip

When the spatial coupling is incorporated into the model, the system exhibits
a rich variety of responses including the uniform state, $\mathbf{m}(z,t)=%
\mathbf{\hat{x}}$, inside the Arnold tongue, subharmonic patterns
predominantly located over the tongue, as well as domain walls, single
solitons, and double solitons. A comprehensive summary of the single-soliton
regime, which arises right under the Arnold tongue, is displayed in Ref.
\cite{Urzagasti2013}. For small values of the modulation drive $h_{0}$, the
soliton shows the usual behavior with subharmonic oscillations. However, for
higher values of the drive, starting from $h_{0}\approx 0.44$, the solitons
exhibit features typical for breathers \cite{Kibler_peregrine_soliton}, such
as the emission of low-amplitude dispersive waves and the oscillation amplitude
that varies on a long timescale. The distinction between the standard and
breather solitons in the space of parameters is revealed by the numerical
analysis based on the fast-Fourier-transform algorithm. It identifies a
well-defined boundary between these dynamical regimes, without any
coexistence region \cite{Urzagasti2013}.

Following the pattern of Ref. \cite{Urzagasti_et_al_JAppPhys_2012}, we here
intend to perform the analysis of the double-soliton response for different
values of the modulation parameters in Eqs. (\ref{eq_h}) and (\ref%
{eq_detuning}), scanning the $(\nu ,h_{0})$ plane. The other parameters are
fixed as in Ref. \cite{Urzagasti_et_al_JAppPhys_2012}, \textit{viz}., $%
\alpha =0.015$, $\beta =20$ and $h_{c}=3$, and the wire's length is chosen
as $2L=250$ (these are values relevant to the respective experimental
setup). As shown in Fig.\ \ref{fig_double_soliton}, the double solitons may
exhibit either the usual (panel (b)) or breathing (panels (c,d)) behavior,
depending of the values of $\nu $ and $h_{0}$. Similar to single-breather
states, double breathing solitons emit dispersive waves. In some cases
(panels (b) and (c)), amplitudes of the bound solitons remain identical in
the course of the long evolution, i.e., the response exhibits symmetry with
respect to the midpoint between the solitons. However, in other cases (panel
(d)), the symmetry is spontaneously broken, and the amplitudes of the bound
solitons may differ significantly. The asymmetric double breathers also
exhibit a slow steady drift, not found in the other single- and
double-soliton regimes, on top of random fluctuations of their positions.
Naturally, the one-sided drift of the breather is made possible by its
spontaneously emerging asymmetry.

\section{The numerical framework}

\label{section_Numerical_framework}

We simulated Eq.\ (\ref{eq_LLG}) by means of FORTRAN 90, using the
fifth-order, double-precision, variable-step Runge-Kutta method, with
monitoring of the local truncation error \cite{NumericalRecipes}. Equation (%
\ref{eq_LLG}) is supplemented by the Neumann's boundary conditions imposed
at edges of the integration domain, \textit{viz}., $\partial \mathbf{m}%
/\partial z=0$ at $z=\pm L$. The accuracy or tolerance level was adopted at
the level of $10^{-5}\%$. The spatial discretization was implemented with $%
dz=1/6\approx 0.17$, in the domain of size $-125<z<+125$. The accuracy of
the numerical scheme was earlier established in Ref. \cite%
{Urzagasti2014,Urzagasti2013}. The simulations were run for different values
of parameters $(\nu ,h_{0})$ in Eqs. (\ref{eq_h}) and (\ref{eq_detuning}),
with $\nu $ and $h_{0}$ ranging from $-0.7$ to $0$, and from $0.3$ to $0.9$,
respectively. For both these parameters, the variation step was $0.002$,
except inside the region of the standard single soliton, where the step was $%
0.01$ (five times coarser), as dynamical indicators of this regime are well
known \cite{Urzagasti2014, Urzagasti2013}. The simulations were run in
parallel using MPI FORTRAN. To assess the reproducibility of the model and
the reliability of the algorithm, simulations were also run using Python
3.11, accelerated with Numba 0.56.

The initial conditions were set symmetrically with respect to $z=0$,
approximately corresponding to two solitons separated by distance $\Delta
z=7 $, while keeping the unitary field magnitude ($|\mathbf{m}|\equiv 1$)
everywhere along the wire:

\begin{align}
m_y(z,t=0) &= \dfrac{1}{2}\big(\sech(z+3.5)+\sech(z-3.5)\big),  \notag \\
m_z(z,t=0) &= 0.95\,m_y,  \notag \\
m_x(z,t=0) &= \sqrt{1-m^2_y-m^2_z}.  \label{eq_init_cond_x} 
\end{align}

The timeline of each simulation was divided into three different stages. The
first one is the transient stage, when the system starts its evolution from
input (\ref{eq_init_cond_x}). The sampling timestep adopted for this stage
was $dt=5^{3}/2^{6}\approx 1.95$, and its duration was $\Delta t=5\times
10^{3}\approx 2560\times dt$, after which we can safely assume that the
system has reached a steady state. During the next, steady-state, stage a
number of dynamical indicators were computed to gather relevant information
about the system, including its response type. This stage had duration $%
\Delta t=10^{3}$, while the sampling timestep was adopted as $%
dt=5^{3}/2^{14}\approx 0.0076\approx \Delta t/131600$, being much shorter
than the modulation period of the AC magnetic field, see Eq. (\ref{eq_h}).
Data for the computation of the largest Lyapunov exponent (LLE) for the
dynamical regimes were collected in the course of the longer third stage,
with duration $\Delta t=14\times 10^{3}$ and sampling time step $%
dt=5^{3}/2^{7}\approx 0.98\approx \Delta /14300$.

\section{Dynamical indicators}

\label{section_Dynamical_indicators}

Addressing the last part of the transient stage of the simulations, \textit{%
viz}., $t\in \lbrack 4000,5000]$, spatially uniform solutions and
fast-moving localized modes were excluded from the analysis. Those states
are irrelevant for the present work, as they are not related to the single-
and double-soliton regimes, which are the target of the consideration. To
distinguish the irrelevant states from the relevant ones, the coordinate of
the center of mass (CM) of the soliton (or solitons) potentially appearing
in the magnetization field was calculated as

\begin{figure}[tbp]
\centering
\begin{subfigure}{0.485\textwidth}
        \centering
        \includegraphics[width=\linewidth]{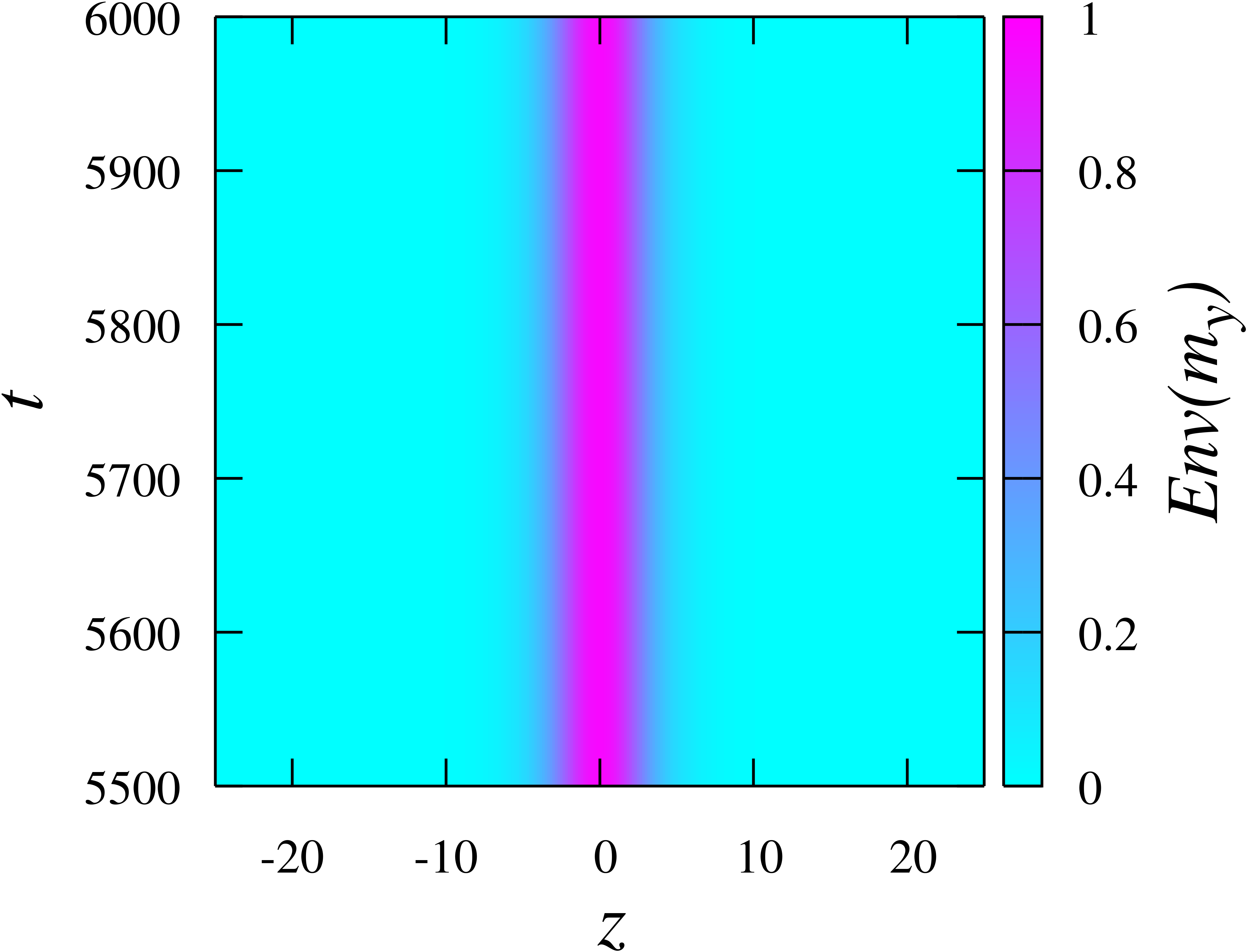}
        \caption{\centering $(\nu,h_0)=(-0.40,0.40)$ Standard soliton}\label{fig_envelopes_ssol}
    \end{subfigure}
\enspace
\begin{subfigure}{0.485\textwidth}
        \includegraphics[width=\linewidth]{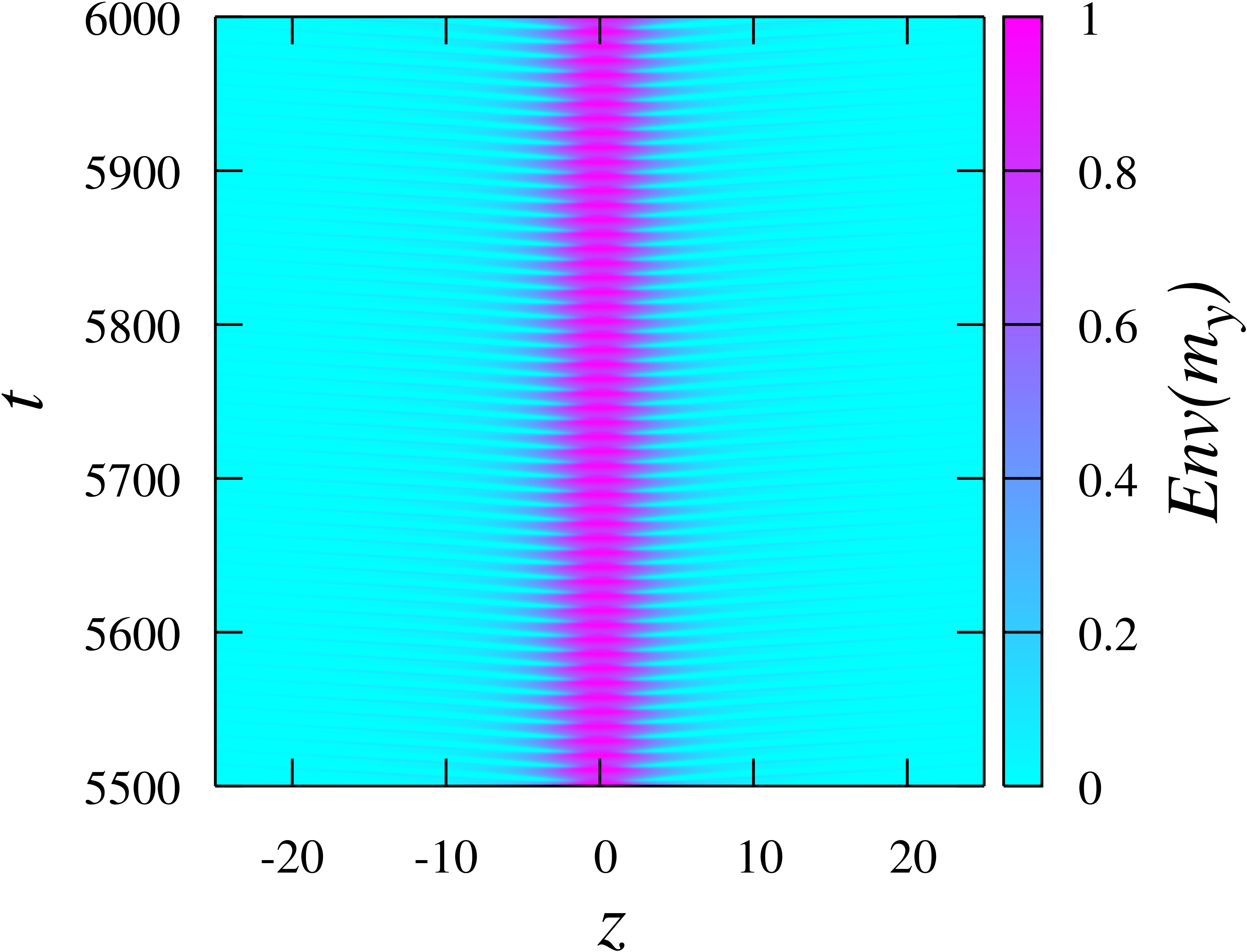}
        \caption{\centering $(\nu,h_0)=(-0.45,0.55)$ Breather soliton}\label{fig_envelopes_bsol}
    \end{subfigure}
\vskip\baselineskip
\begin{subfigure}{0.485\textwidth}
        \centering
        \includegraphics[width=\linewidth]{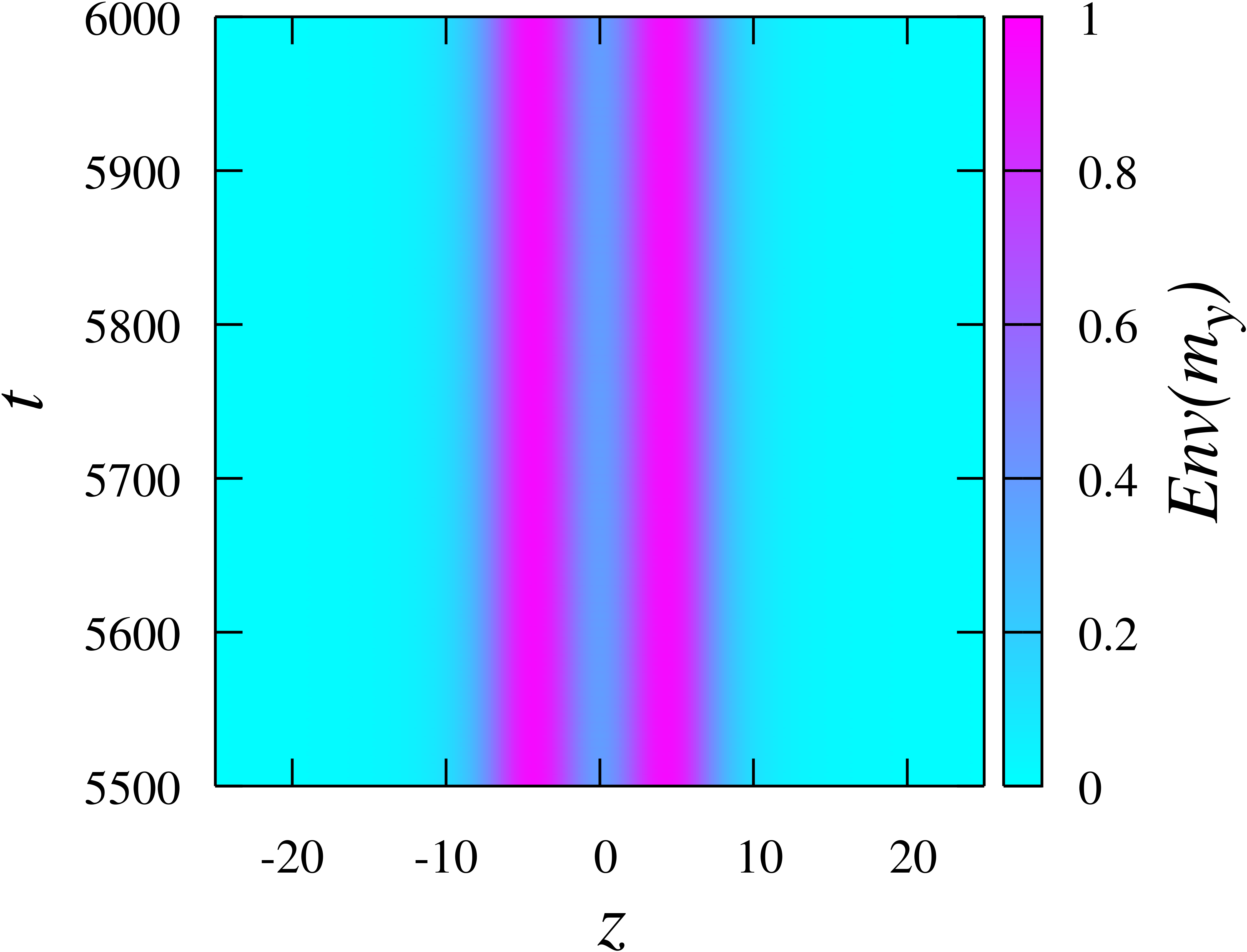}
        \caption{\centering $(\nu,h_0)=(-0.30,0.55)$ Double standard soliton}\label{fig_envelopes_dsol}
    \end{subfigure}
\enspace
\begin{subfigure}{0.485\textwidth}
    	\vspace*{0.5cm}
        \includegraphics[width=\linewidth]{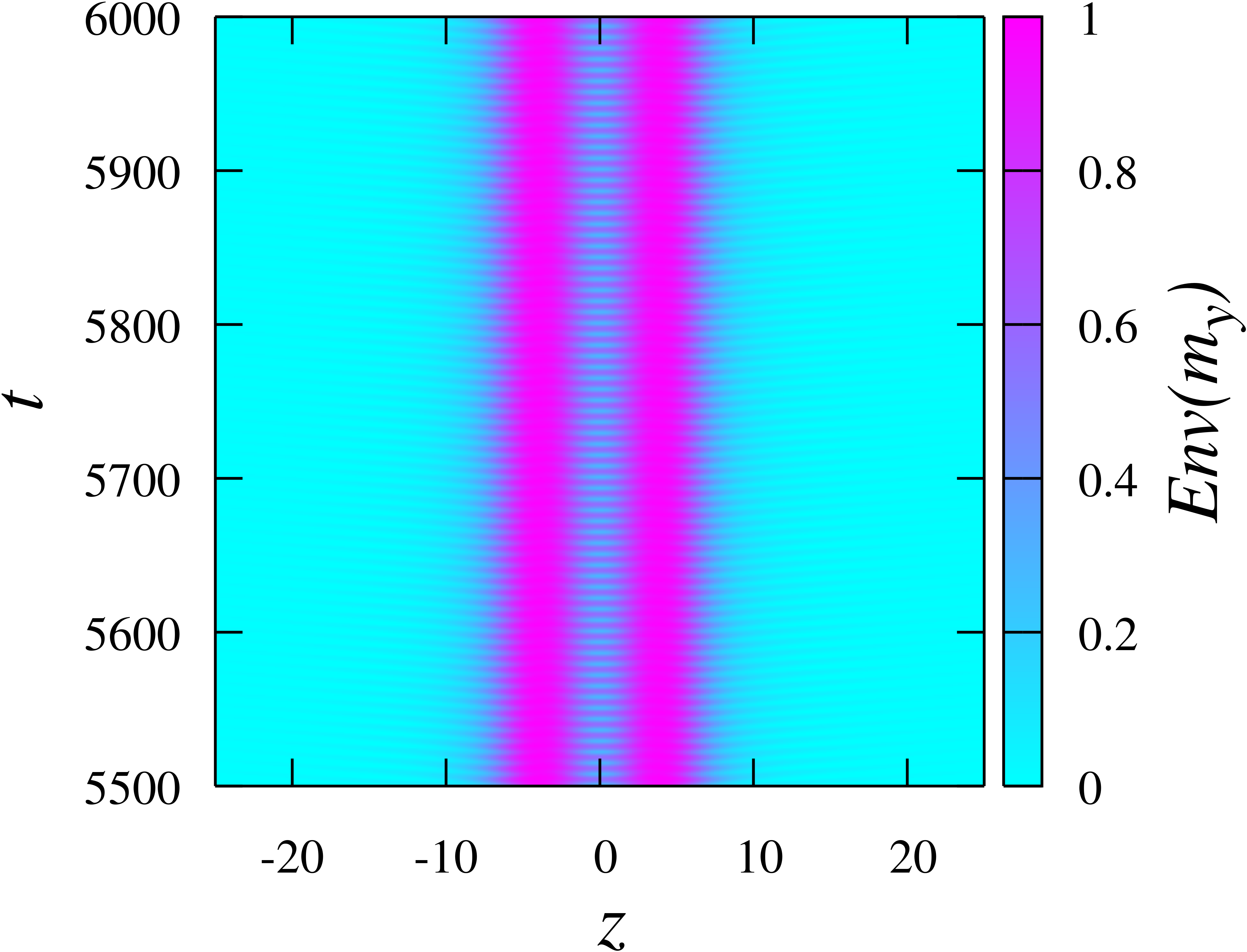}
        \caption{\centering $(\nu,h_0)=(-0.37,0.61)$ Double breather symmetric soliton}\label{fig_envelopes_dbsol}
    \end{subfigure}
\vskip\baselineskip
\begin{subfigure}{0.485\textwidth}
        \centering
        \includegraphics[width=\linewidth]{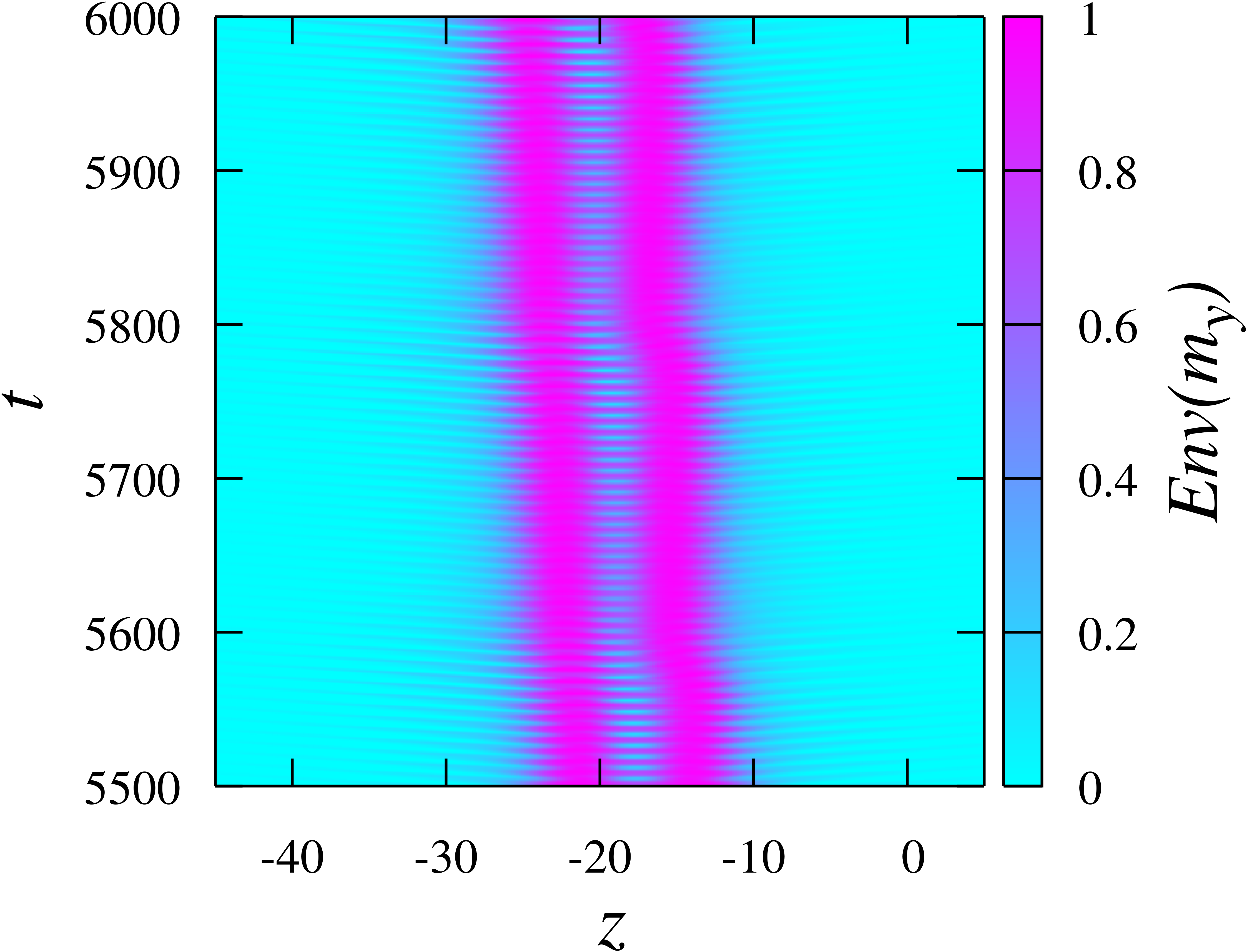}
        \caption{\centering $(\nu,h_0)=(-0.35,0.65)$ Double breather asymmetric soliton with drift}\label{fig_envelopes_dbdsol}
    \end{subfigure}
\enspace
\begin{subfigure}{0.485\textwidth}
    	\vspace*{0.5cm}
        \includegraphics[width=\linewidth]{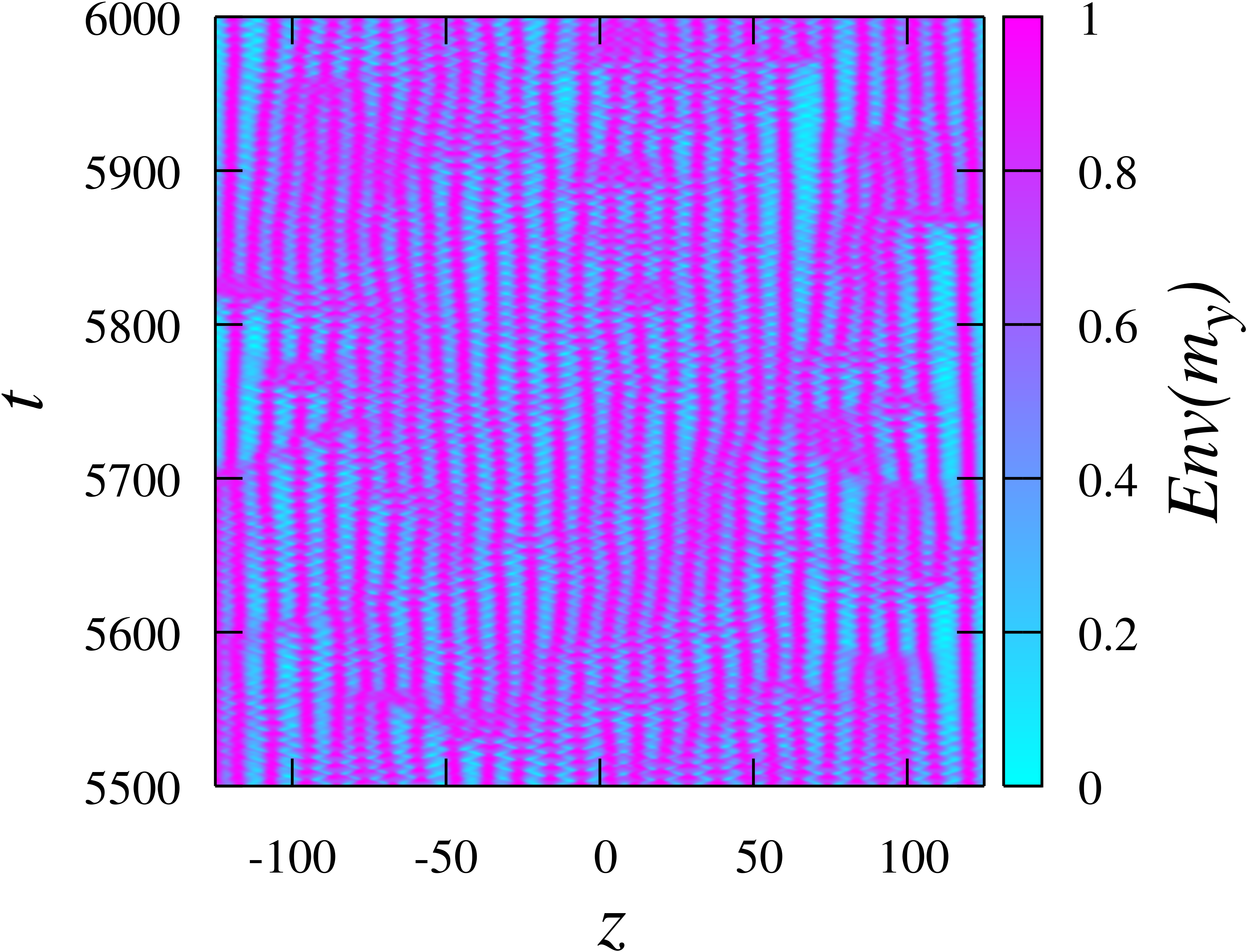}
        \caption{\centering $(\nu,h_0)=(-0.35,0.65)$ Subharmonic Pattern \qquad \qquad}\label{fig_envelopes_pattsol}
    \end{subfigure}
\vskip\baselineskip
\caption{(color online) Envelopes of different responses of the magnetic
wire, defined as per Eq. (\protect\ref{eq_mean_my}) and plotted as $%
m_{y}(z,t_{i})$, with $t_{i}$ being time moments at which the spatial
average of $m_{y}(z,t)$ has a local maximum. Therefore, the color maps
capture the dynamics only for the long timescale.}
\label{fig_envelopes}
\end{figure}
\unskip

\begin{equation}
z_{\mathrm{CM}}(t)=\dfrac{\int_{-L}^{+L}z\cdot m_{y}^{2}(z,t)\,dz}{%
\int_{-L}^{+L}m_{y}^{2}(z,t)\,dz}.  \label{eq_z_cm}
\end{equation}%
We focused on the $m_{y}$ component of the magnetization field, as $m_{y}=0$
in the uniform state, while solitons give rise to a local excitation of $%
m_{y}$. The CM mean velocity of the drifting solitons was found by fitting
the time series $z_{\mathrm{cm}}(t)$ to a linear dependence, and taking its
slope. Modes with the drift velocity exceeding $0.005$ are considered as
fast-moving ones and dismissed from the subsequent analysis due to artifacts
expected from their collisions with the boundaries of the simulation box. This
problem will be eliminated if a ring-shaped wire is considered, with
periodic boundary conditions, which will be the subject of a separate work.

Additionally, the following integral characteristic was computed,

\begin{equation}
I_{D}=\int_{4000}^{5000}\bigg[\dfrac{1}{2L}\int_{-L}^{+L}\left(
m_{y}(z,t)\right) ^{2}dz\bigg]^{2}dt.  \label{eq_id}
\end{equation}%
It provides a measure of the departure of the given state from the uniform
one, which has $I_{D}\equiv 0$. Subsequently, $I_{D}$ can be used to roughly
estimate what share of the entire wire is occupied by the solitons or
patterns. States with $I_{D}<10^{-10}$ are assumed to be uniform ones, while
those with $I_{D}>10^{-2}$ are considered as patterns.

While considering the steady-state stage, in the course of interval $t\in
\lbrack 5000,6000]$, attention was again focused the $m_{y}$ component. The
short-timescale dynamics of $\mathbf{m}$ in response to the external field
modulation was eliminated by computing the envelope of $m_{y}$. The
evolution of the envelope is represented by values $m_{y}(z,t_{i})$, where $%
t_{i}$ are time points maximizing the space average of $m_{y}$,

\begin{equation}
\langle m_{y}\rangle (t)=\frac{1}{2L}\int_{-L}^{+L}m_{y}(z,t)dz.
\label{eq_mean_my}
\end{equation}

Thus, only variations of the response amplitude on the long timescale are
taken into account. Figure \ref{fig_envelopes} shows examples of the
envelope evolution for different responses observed at the steady-state
stage for different values of the control parameters, $(\nu ,h_{0})$. Panels
(a) and (b) show the standard single-soliton and breather regimes, which are
similar to those previously reported in Ref. \cite%
{Urzagasti_et_al_JAppPhys_2012}. Panel (c) shows an example of the standard
two-soliton pair. In that case, individual solitons in the pair keep the
amplitude and width similar to those of the single-soliton states.

On the other hand, panels (d) and (e) in Fig. \ref{fig_envelopes} display
novel double-breather soliton solutions, whose amplitudes oscillate on the
long timescale, unlike the non-breathing solitons, which remain completely
stationary. Both single- and double-breather solitons emit radiation.
Furthermore, the double breather in panel (e) in Fig. \ref{fig_envelopes} is
drifting away from its initial position, which is closely related to the
noticeable breaking of its inner symmetry. The drift is exhibited by all
double-breather solitons (it is not visible in panel (d), as the timescale
is not long enough for that). Indeed, the asymmetric emission of radiation
resulting from the spontaneously emerging asymmetry in the structure of the
soliton complexes gives rise to unequal left and right recoil forces, which
drives the drift. Finally, panel (f) in Fig. \ref{fig_envelopes} shows an
example of the dynamical regime in which the entire wire is occupied by an
apparently randomized subharmonic pattern.

\begin{figure}[tbp]
\centering
\begin{subfigure}{0.48\textwidth}
        \centering
        \includegraphics[width=\linewidth]{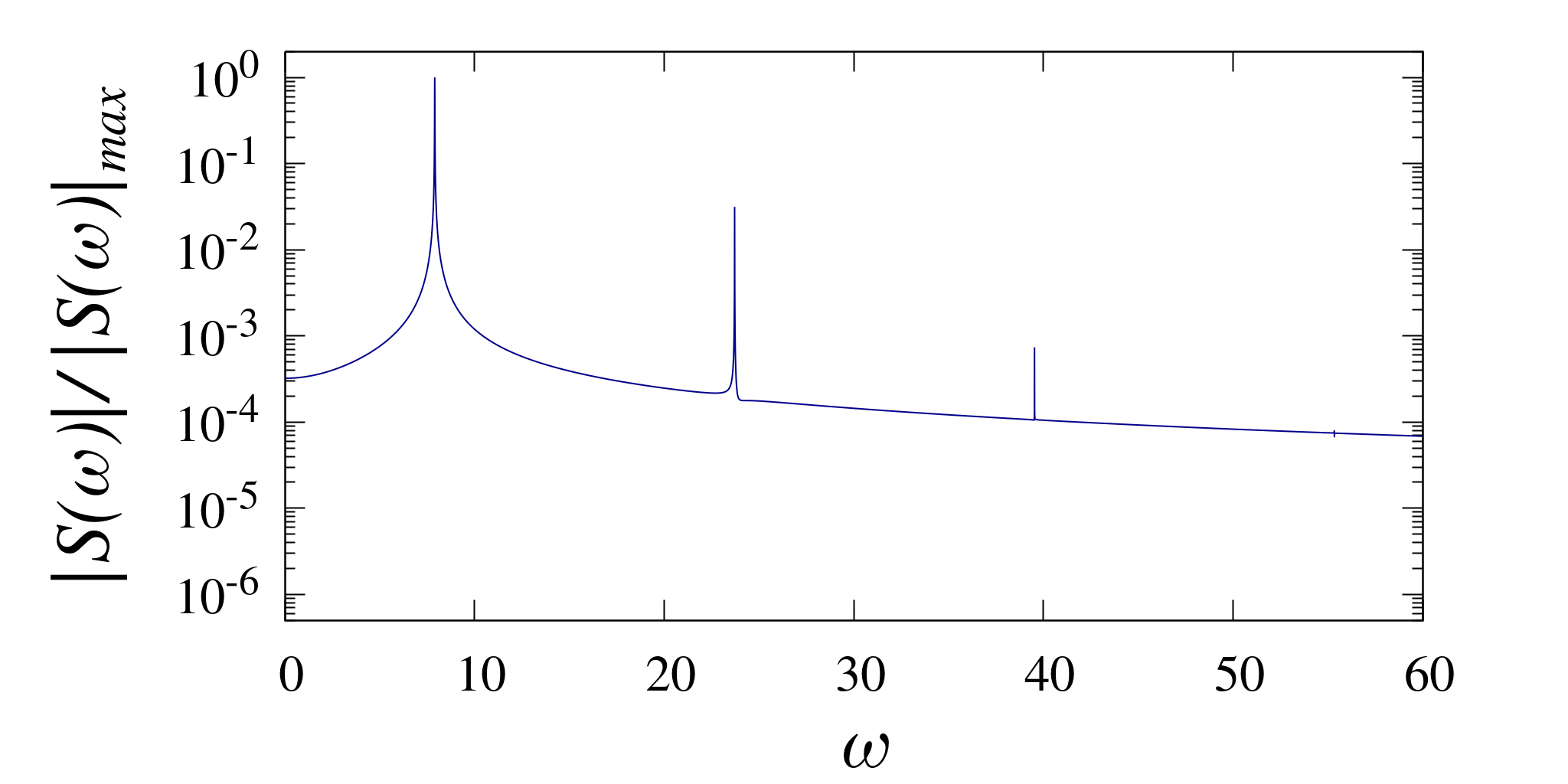} \\
        \includegraphics[width=\linewidth]{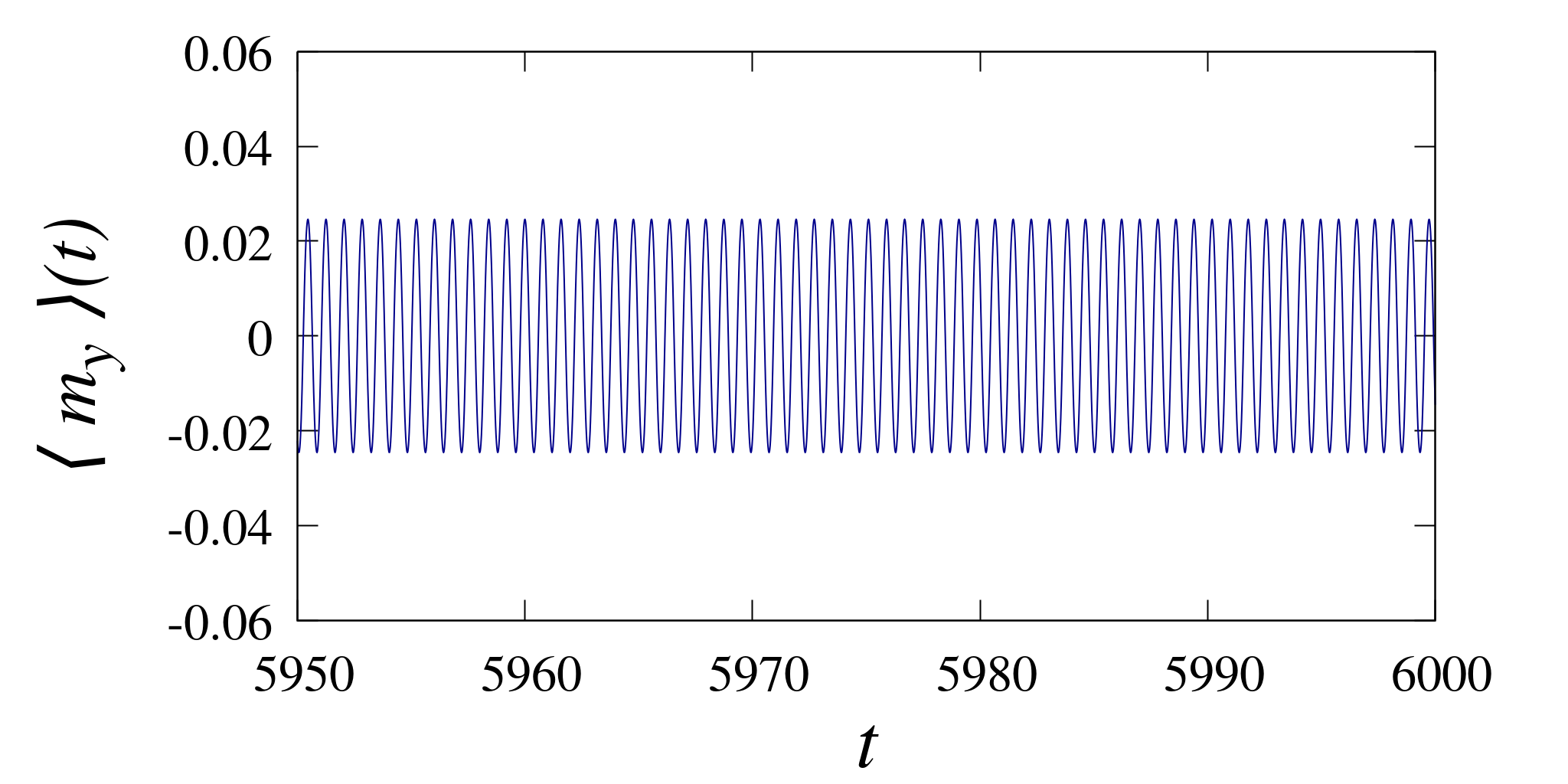}
        \caption{\centering $(\nu,h_0)=(-0.40,0.40)$ Standard soliton}\label{fig_Im_fft_ssol}
    \end{subfigure}
\begin{subfigure}{0.48\textwidth}
        \centering
        \includegraphics[width=\linewidth]{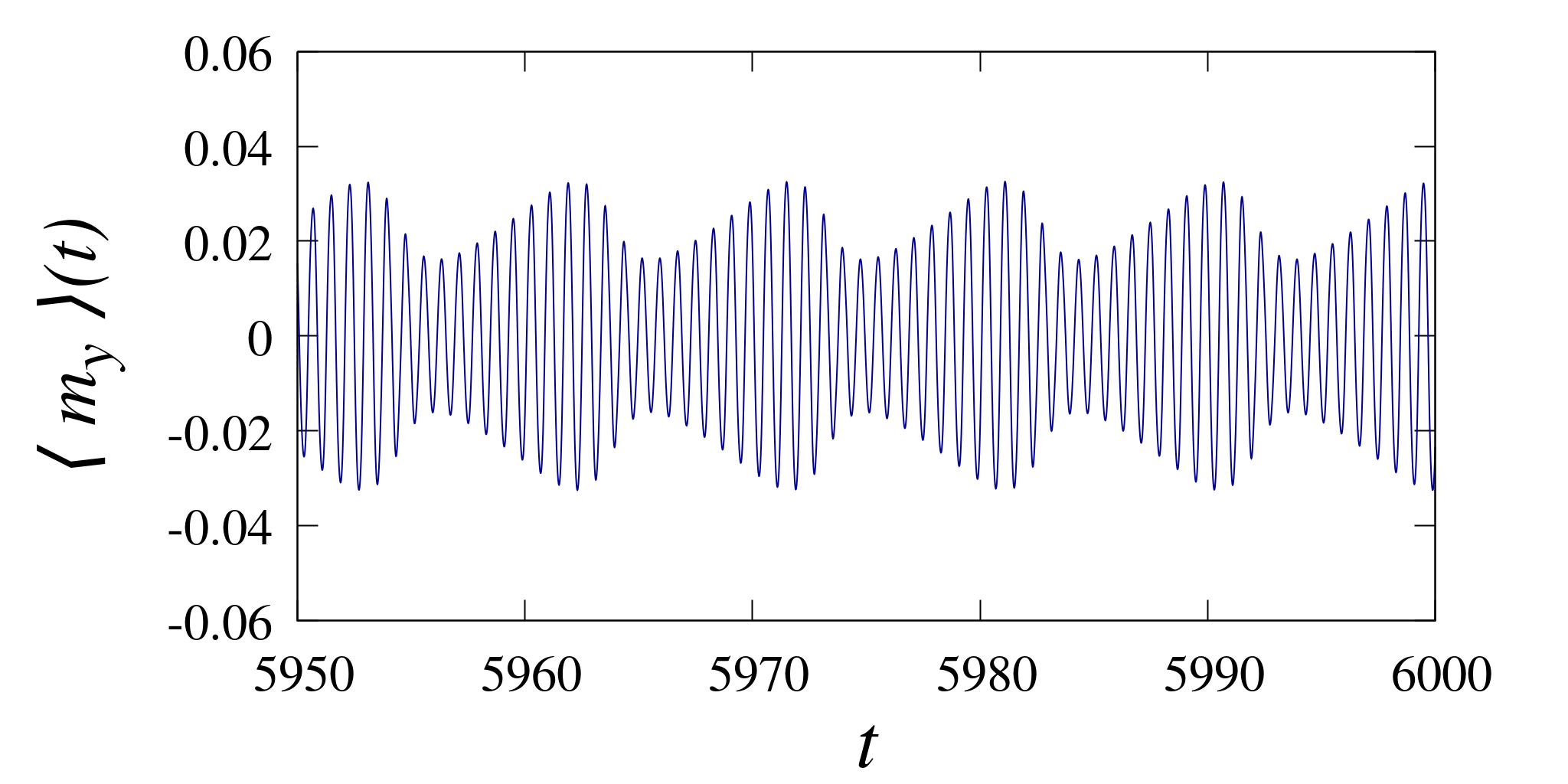} \\
        \includegraphics[width=\linewidth]{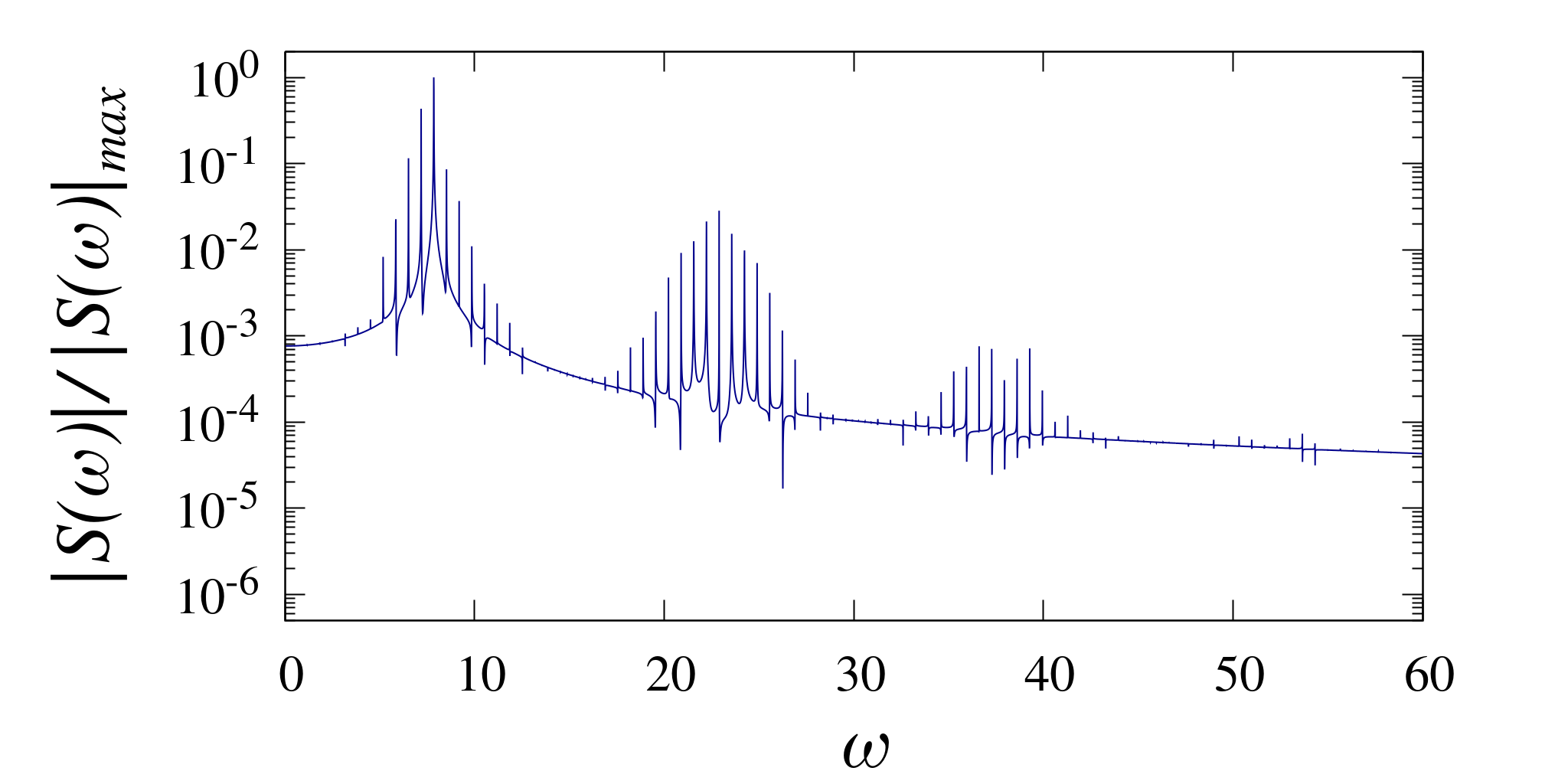}
        \caption{\centering $(\nu,h_0)=(-0.45,0.55)$ Breather soliton}\label{fig_Im_fft_bsol}
    \end{subfigure}
\vskip\baselineskip
\begin{subfigure}{0.48\textwidth}
        \centering
        \includegraphics[width=\linewidth]{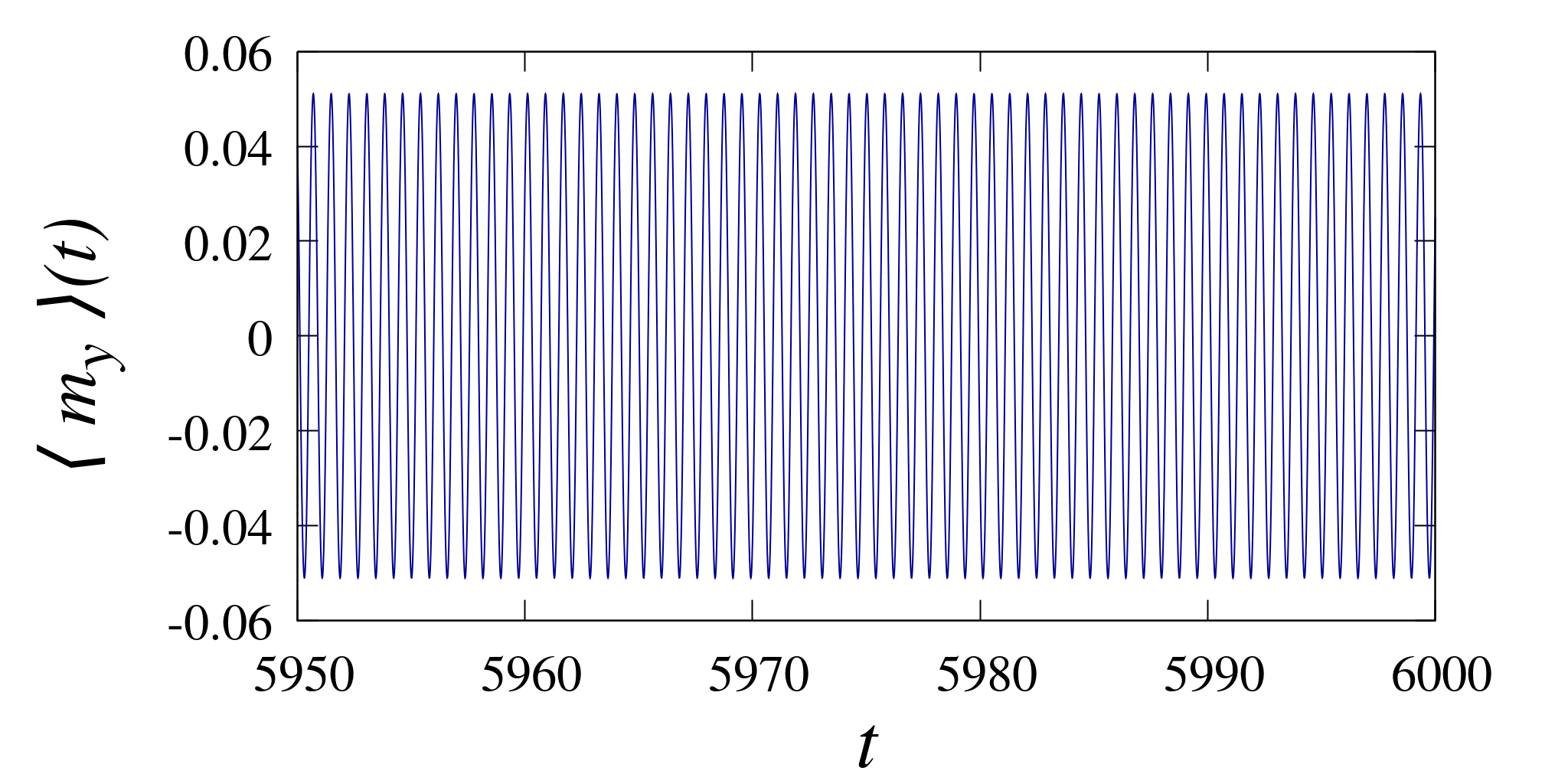} \\
        \includegraphics[width=\linewidth]{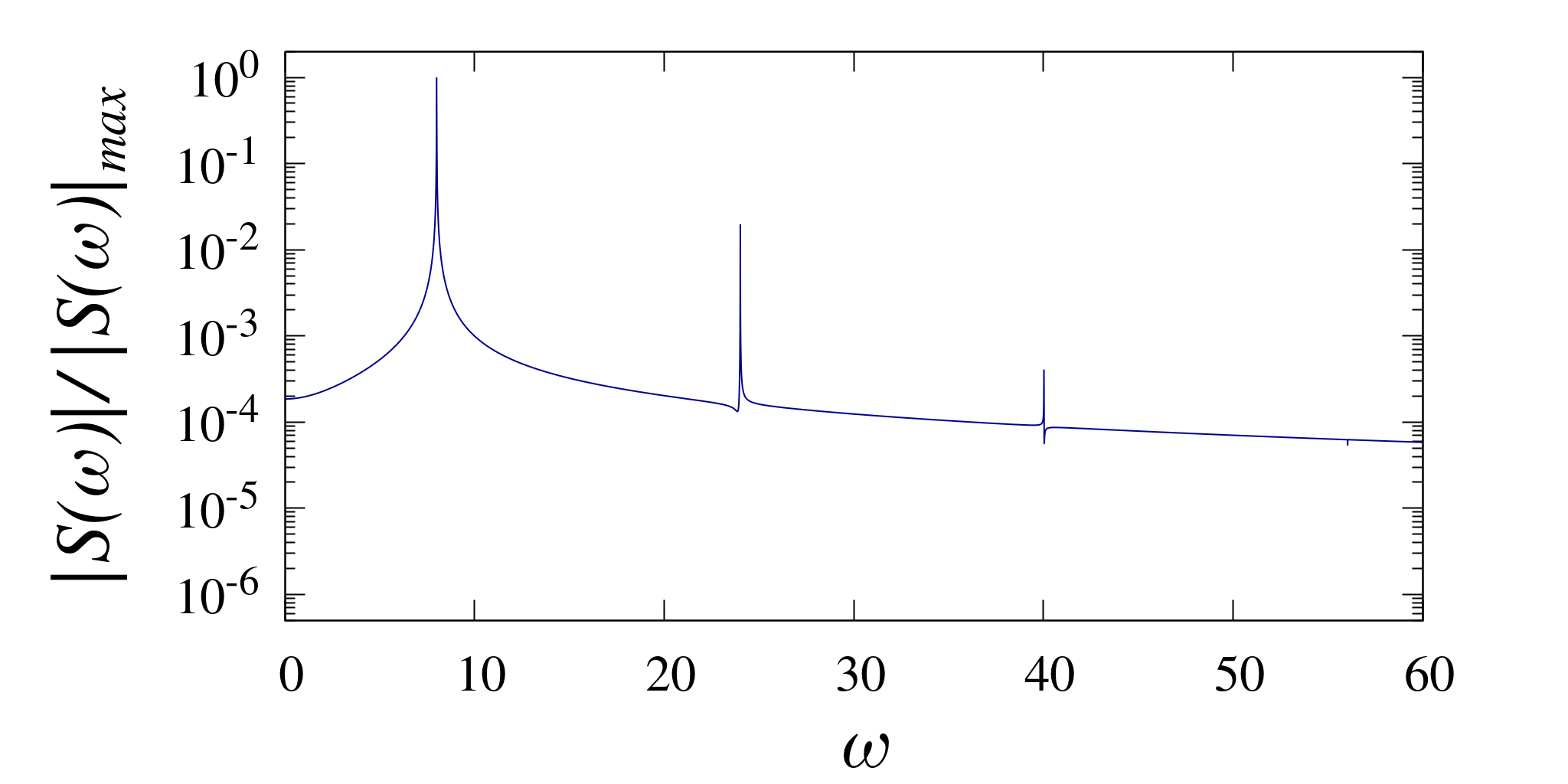}
        \caption*{\textbf{3.c}\centering $(\nu,h_0)=(-0.30,0.55)$ Double standard soliton}\label{fig_Im_fft_dsol}
    \end{subfigure}
\begin{subfigure}{0.48\textwidth}
        \centering
        \includegraphics[width=\linewidth]{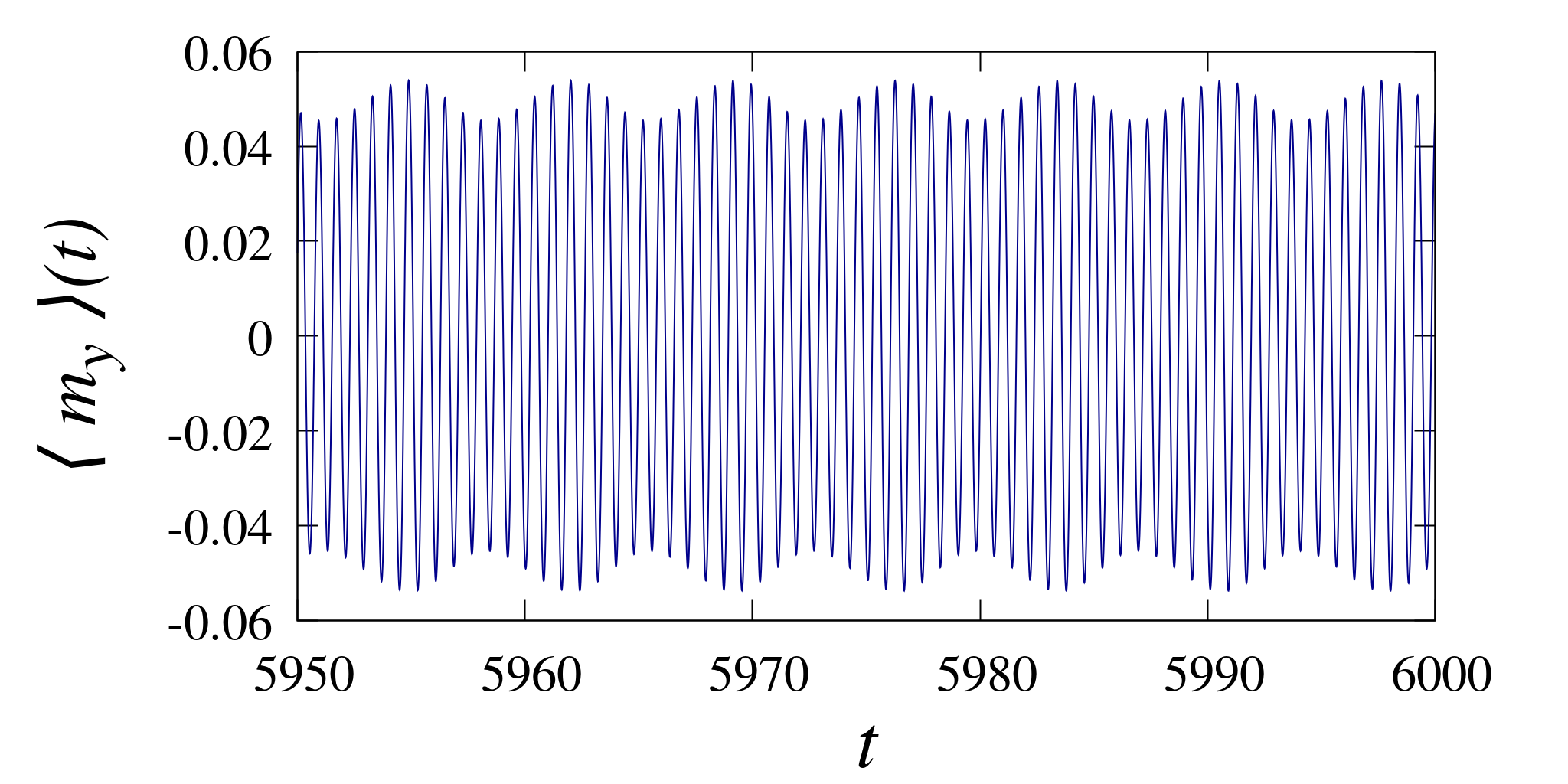} \\
        \includegraphics[width=\linewidth]{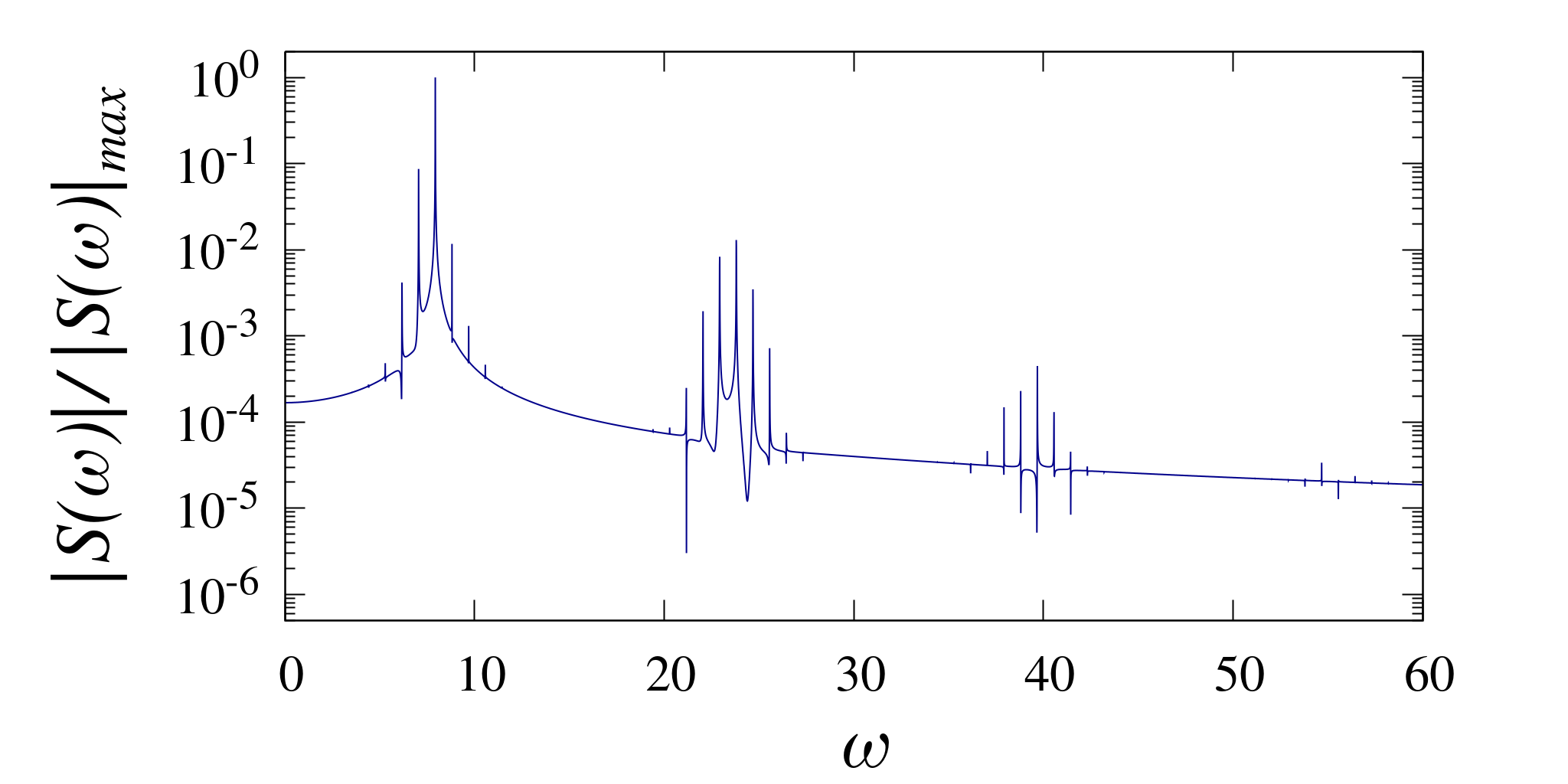}
        \caption*{\textbf{3.d}\centering $(\nu,h_0)=(-0.37,0.61)$ Double breather soliton}\label{fig_Im_fft_dbsol}
    \end{subfigure}
\vfill
\caption{Plots of the spatial average of $m_{y}(x,t),$ defined as per Eq. (%
\protect\ref{eq_mean_my}) and its normalized power spectral density (PSD)
for the soliton regimes displayed in Fig. \protect\ref{fig_envelopes}. The
standard soliton states exhibit the spatial average oscillating with a
constant amplitude, and, accordingly, PSD with well-defined peaks at odd
multiples of the main oscillation frequency, $\protect\omega =\Omega /2$,
which is imposed by the parametric drive in Eq. (\protect\ref{eq_h}). In
breather solitons, the oscillation amplitude of the spatial average varies
on the long timescale, which translates into PSD with multiple frequency
peaks.}
\label{fig_Im_fft}
\end{figure}
\unskip

In the steady stage, the CM position of the envelope was defined in the same
way as in Eq. (\ref{eq_z_cm}), with the difference that here we consider
only time moments $t_{i}$ at which $\langle m_{y}\rangle (t)$ has a maximum.
The average CM velocity is then estimated, as said above, by means of the
linear fit to $z_{\mathrm{CM}}(t_{i})$. To provide a simple criterion
distinguishing single and double solitons in the response, the time average
of the centered envelope,%
\begin{equation}
\bar{m}_{\mathrm{env}}(z)=\frac{1}{n}\sum_{i=1}^{n}m_{y}(z-z_{\mathrm{CM}%
}(t_{i}),t_{i})
\end{equation}%
was computed. In the framework of the time-averaging procedure, the
radiation waves emitted by the breathers were flattened, and the number of
solitons was identified by directly counting the number of local-maximum
points in the $\langle m_{y}\rangle (z)$ profile.

Regardless of the number of solitons, the distinction between fundamental
ones and breathers was determined by computing the normalized power spectral
density (PSD) of $\langle m_{y}\rangle (t)$, i.e., the squared absolute
value of the complex fast Fourier transform, normalized to its maximum.\
Figure \ref{fig_Im_fft} presents plots of both $\langle m_{y}\rangle (t)$
and its PSD corresponding to the same responses which are displayed in Fig. %
\ref{fig_envelopes}. In the usual regimes (panels (a) and (c)), the
soliton's oscillation amplitude remains constant, and the PSD exhibits a
global peak at $\omega =\Omega /2$ (i.e., half the driving frequency in the
parametrically forced equation (\ref{eq_h})), as the response is
subharmonic, and it also exhibits lower peaks at odd multiples of $\Omega /2$%
. If the breathing is present (panels (b) and (d)), the soliton's size
varies on the long timescale, and PSD exhibits a multitude of low peaks
agglomerated around the odd multiples of $\Omega /2$, representing a large
set of oscillation modes. The simple peak count makes it possible to
conclude if the breathing is present or not in the dynamical regime. To this
end, only the peaks at frequencies $\omega <3\Omega $ are considered, as
peaks beyond this range are extremely weak.

The microscopically defined magnetic free energy $Q$ of the system is the
sum of the exchange energy, anisotropy energy, and the energy of the
interaction with the external magnetic field \cite{NL_Magnetization}, i.e.,

\begin{equation}
Q=\dfrac{1}{2L}\int_{-L}^{+L}\bigg(\dfrac{1}{2}(\partial _{z}\mathbf{m})^{2}-%
\dfrac{\beta }{2}(1-m_{z}^{2})-\mathbf{h}\cdot \mathbf{m}\bigg)\,dz.
\label{eq_magener}
\end{equation}

The last dynamical indicator employed in our analysis is the largest
Lyapunov exponent (LLE), $\lambda _{\max }$, which provides a deeper
understanding of the dynamical system. In particular, $\lambda _{\max }$
determines the average exponential rate of divergence or convergence of
neighboring orbits in the system's phase space. Thus, condition $\lambda
_{\max }>0$ delineates a region of chaotic dynamics, suggesting rapid loss
of the system's predictability, which is the hallmark of dynamical chaos.
The magnitude of positive $\lambda _{\max }$ defines the timescale on which
the dynamics become unpredictable. Conversely, exponential convergence,
corresponding to $\lambda _{\max }<0$, indicates stability of periodic
orbits. In the neutral case, $\lambda _{\max }\approx 0$, the system is
defined as a marginally stable or quasi-periodic one.

The LLE can be calculated as \cite{WOLF1985},

\begin{equation}
\lambda _{\max }=\lim_{t\rightarrow \infty }\dfrac{1}{t}\ln \dfrac{||\delta
\mathbf{m}(z,t)||}{||\delta \mathbf{m}(z,t_{0})||},  \label{eq_LLE}
\end{equation}%
where $||\bullet ||\equiv (\int_{-L}^{+L}|\bullet |^{2}\,dz)^{1/2}$. The
vector function $\delta \mathbf{m}$ in Eq. (\ref{eq_LLE}), which represents
the distance in the phase space between close orbits, obeys the linear
equation,

\begin{equation}
\dfrac{\partial (\delta \mathbf{m})}{\partial t}=\mathbf{\overline{J}}\cdot
\delta \mathbf{m},  \label{J}
\end{equation}%
where $\mathbf{\overline{J}}$ is the Jacobian matrix of Eq. (\ref{eq_LLG}).
In our numerical analysis, $\lambda _{\max }$ was calculated from $%
t_{0}=6\times 10^{3}$ up to $t_{\max }=2\times 10^{4}$. At each time step, $%
||\delta \mathbf{m}(z,t)||$ was rescaled to restore its initial norm, $%
||\delta \mathbf{m}(z,t_{0})||$. Finally, $\lambda _{\max }$ is evaluated
as the average value between times $t=1.8\times 10^{4}$ and $t=2.0\times
10^{4}$. Let us comment that this method is the most common one used in the literature to characterize chaotic states \cite{sano1985measurement,ramasubramanian2000comparative,rosenstein1993practical,geist1990comparison,
pati2024spiral,mazanik2024hysteresis,bazzani2023performance,field2021quint,zhao2024multiple}.

\section{Results}

\label{section_Results}

\subsection{Existence regions}

\label{subsection_ER}

\begin{figure}[tbp]
\centering
\begin{subfigure}{0.485\textwidth}
        \includegraphics[width=\linewidth]{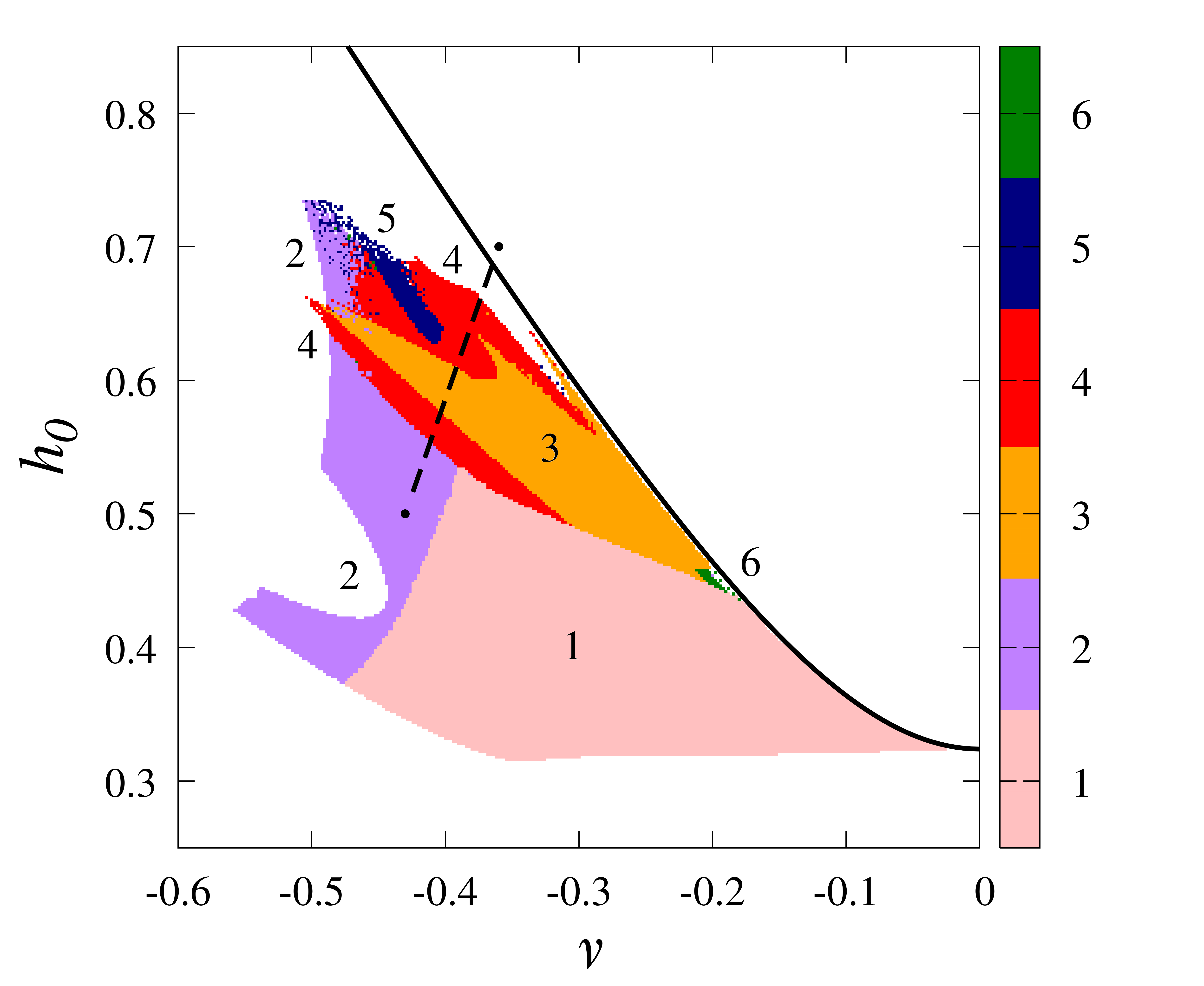}
        \caption{\centering Existence regions }\label{fig_ER}
    \end{subfigure}
\begin{subfigure}{0.485\textwidth}
        \includegraphics[width=\linewidth]{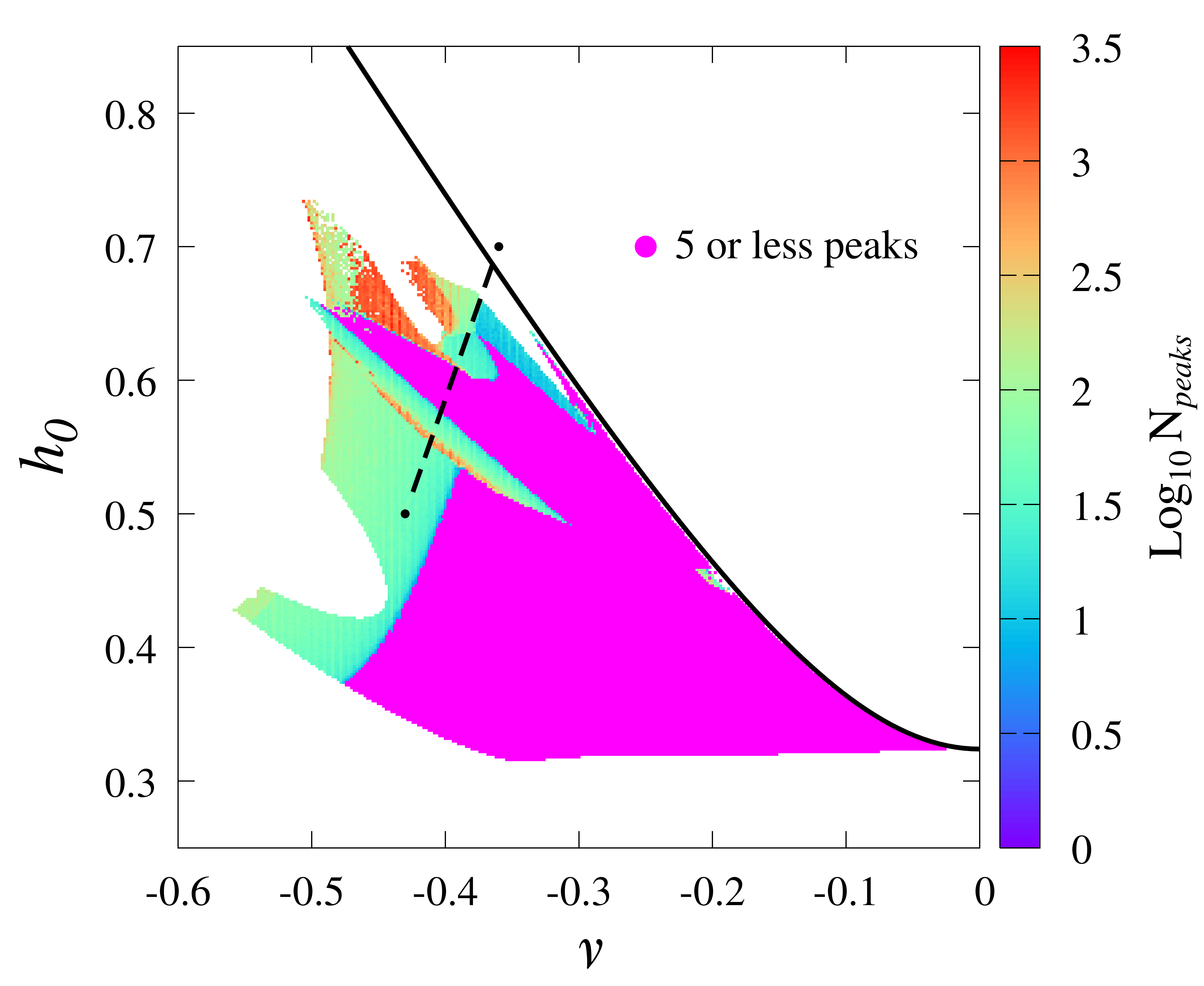}
        \caption{\centering Fast Fourier Transform}\label{fig_fftnvt}
    \end{subfigure}
\caption{(color online) (\textbf{a}) Existence regions in the $(\protect\nu %
,h_{0})$ parameter plane. The regions are found for the following six
states: simple solitons ($1$--pink), breather solitons ($2$--purple), double
solitons ($3$--orange), double breather solitons ($4$--red), fast-moving
localized structures ($5$--blue), and intermittent localized states ($6$%
--green). (\textbf{b}) Number of peaks in the case of $\protect\omega %
=3\Omega $, as produced by the fast Fourier transform. In both panels, the
solid line is the boundary of the first Arnold tongue. The modes belonging
to the dashed line, $h_{0}(\protect\nu )=2.857\protect\nu +1.729$, ranging
from $(\protect\nu ,h_{0})=(-0.43,0.5)$ to $(\protect\nu ,h_{0})=(-0.36,0.7)$%
, are considered by means of the detailed study of the micromagnetic energy
and largest Lyapunov exponent in subsections \protect\ref{subsection_Q} and
\protect\ref{subsection_LLE}, respectively.}
\label{fig_ER_fftnvt}
\end{figure}

Figure \ref{fig_ER_fftnvt}(a) summarizes all localized-state responses in
the $(\nu ,h_{0})$ map for the initial condition given by Eqs. (\ref%
{eq_init_cond_x}). The existence regions lie right under the boundary of the
first Arnold tongue, above which the system exhibits only delocalized
patterns covering the entire spatial domain. Among the existence regions,
the lowest and largest ones are those where the system responds with the
creation of a standard single soliton ($1$--pink) or a single breather
soliton ($2$--purple). Both regions are significantly larger than their
counterparts found in Ref. \cite{Urzagasti2014}, almost doubling their area.
This happens because a different initial condition, in the form of a simple
soliton, was used in that work. Surprisingly enough, such initial condition
falls outside the attraction basin of the single-soliton response for lower
values of $h_{0}$, yet the double-soliton initial condition, defined above
by Eqs. (\ref{eq_init_cond_x}), belongs to the respective basin.

On the other hand, the double-soliton initial condition does produce a
stable double-soliton response above the aforementioned regions, still under
the boundary of the Arnold tongue, between $\nu =-0.5$ and $\nu =0.2$. Three
regions are distinguished here. In the middle one ($3$--orange), the double
soliton is of the usual type, with no breathing, while near the lower-left
border and in the upper part ($4$--red) the double soliton exhibits
breathing (thus this response is found in a disconnected set). Some of
these modes exhibit the drift at a relatively small velocity. The region in
the upper-left corner ($5$--blue) corresponds to the responses that are
identified as fast-drifting localized structures, according to the
measurement of their CM velocity, as outlined at the beginning of section %
\ref{section_Dynamical_indicators}. Here, the system exhibits breathing
double solitons. Drifting at larger velocities, they hit edges of the
spatial domain, during either the transient stage of the steady one. The
simulations reveal that, as a result of hitting the edge, the double soliton
stays stuck near the edge for an uncertain time, and then drifts back. In
the double-soliton regions, the single-soliton initial condition still leads
to the stable single-soliton responses \cite{Urzagasti2014}, which implies
the system's multistability, as the single- and double-soliton regimes
coexist with each other and with the uniform state, $\mathbf{m}=\hat{x}$.

\unskip

\begin{figure}[tbp]
\centering
\begin{subfigure}{0.485\textwidth}
        \centering
        \includegraphics[width=\linewidth]{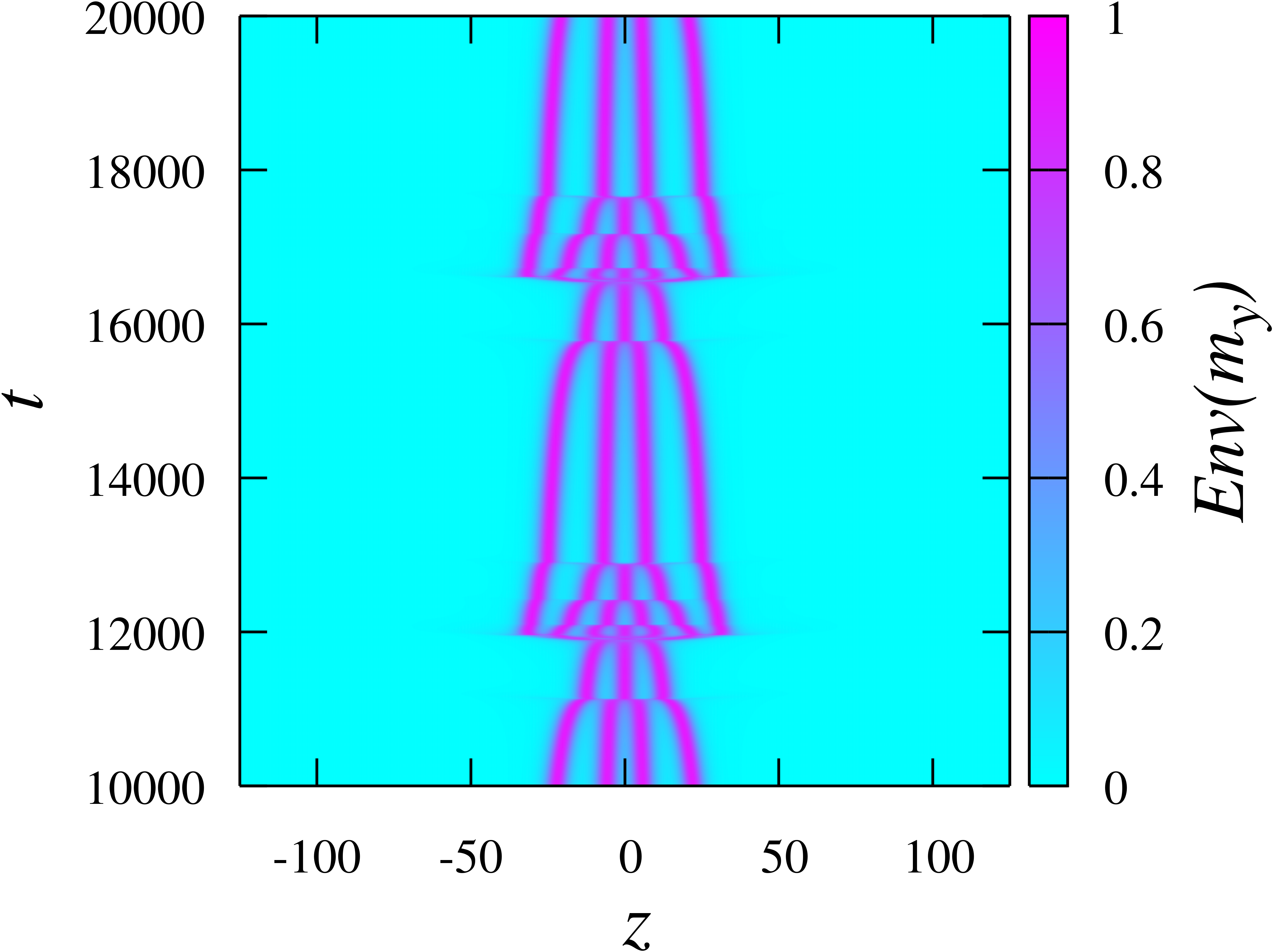}
        \caption{\centering}
    \end{subfigure}
\begin{subfigure}{0.400\textwidth}
        \centering
        \includegraphics[width=\linewidth]{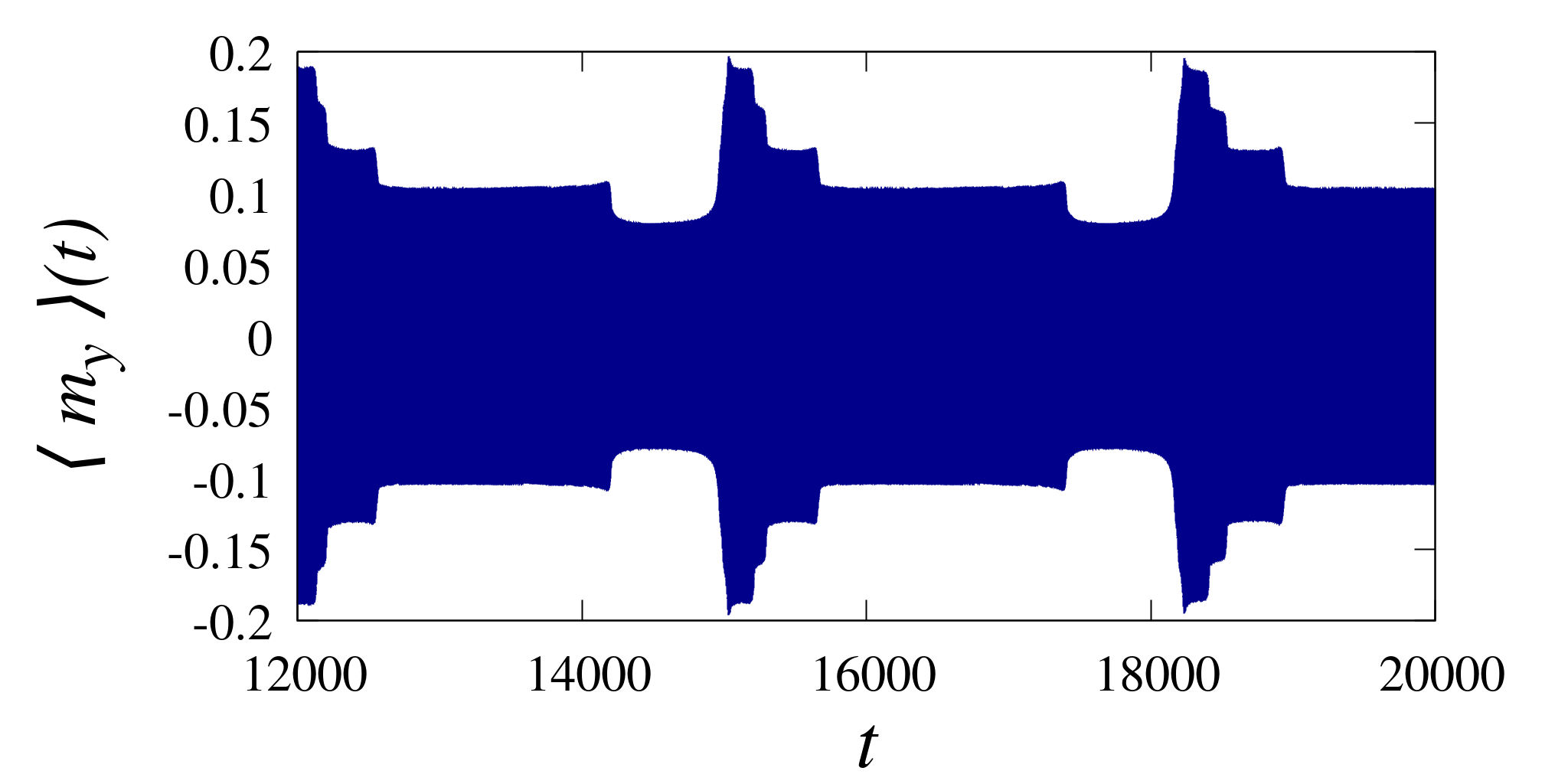} \\
        \includegraphics[width=\linewidth]{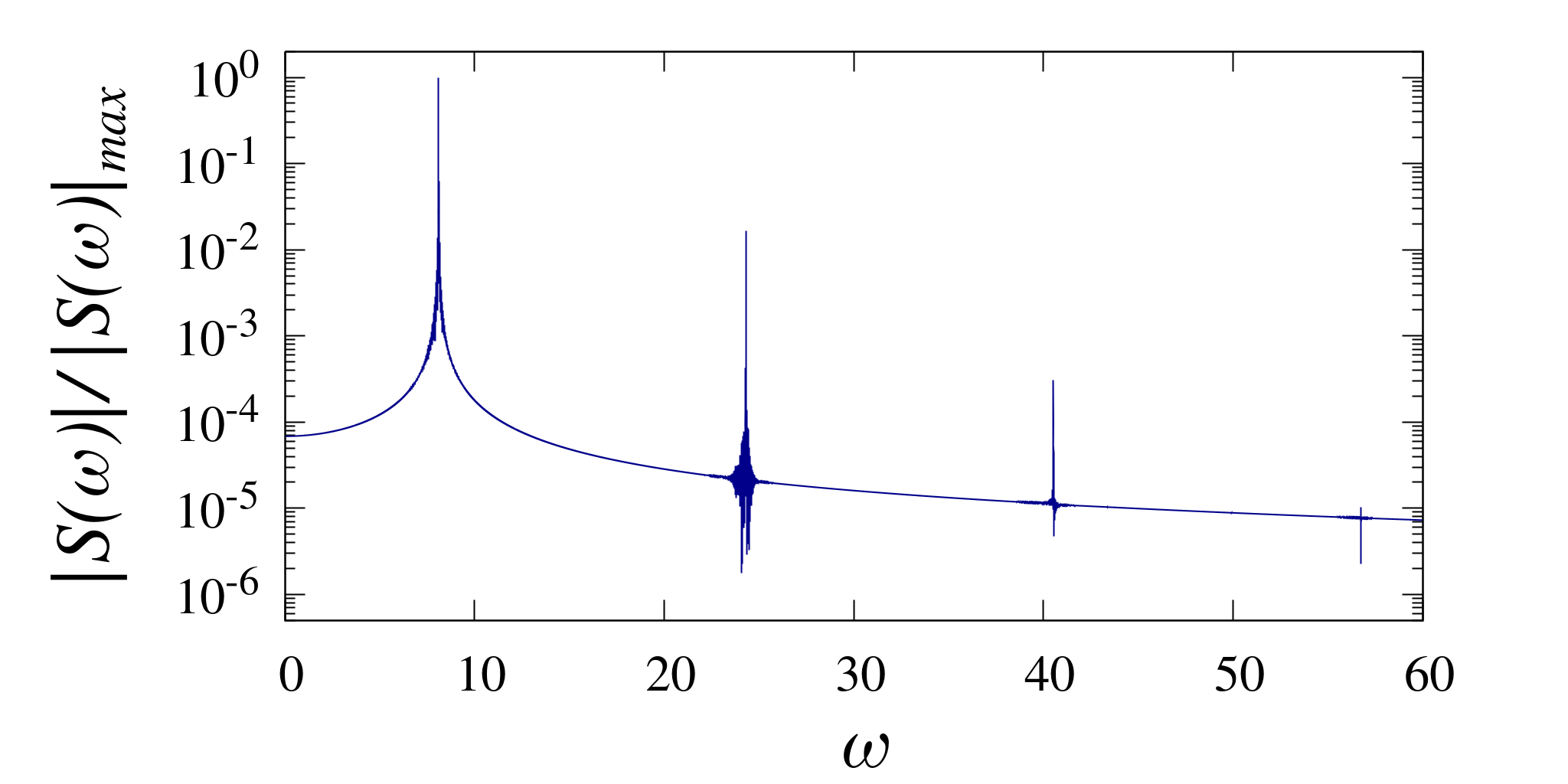}
        \caption{\centering}
    \end{subfigure}
\vskip\baselineskip
\caption{(color online) The response of the system for $(\protect\nu %
,h_{0})=(-0.20,0.45)$ in the form of an intermittent localized state. (%
\textbf{a}) Envelope $m_{y}(z,t_{i})$. (\textbf{b}) The spatial average of $%
m_{y}(x,t)$ and its normalized power spectral density (PSD).}
\label{fig_intermittent}
\end{figure}

An unexpected response was observed in a very small region around $(\nu
,h_{0})=(-0.20,0.45)$, just under the Arnold tongue (region $6$--green in
Fig. \ref{fig_ER_fftnvt}(a)). Here, the system exhibits a localized
intermittent state where a small region in the wire is filled by solitons
whose number varies in time, as they may spontaneously appear, disappear or collide.
Figure \ref{fig_intermittent}(a) illustrates an example of the intermittent
state. The two initially launched solitons collide and merge into a single
soliton, which suddenly splits in a set of seven solitons, that occupy the
region between $z=-50$ and $z=+50$ with a relatively smooth distribution.
Shortly afterwards, the solitons drift towards the center, leading to
disappearance of the central soliton, and thus decreasing the soliton number
to six. Next, two adjacent solitons collide and merge, which further reduces
the number to five. As the solitons keep drifting to the center, the
above-mentioned soliton, produced by the merger, also disappears while two
others, adjacent to it, collide and merge once again. The remaining three
solitons become closer and suddenly split in seven solitons, then repeating the
cycle of the solitons metamorphoses. This dynamical regime is very robust
and regular in its evolution. It is spatially symmetric and does not exhibit
a global drift. The regularity of this regime is also confirmed in Fig. \ref%
{fig_intermittent}(b) by the respective picture of the evolution of the
spatial average $\langle m_{y}\rangle (t)$, as well as by its PSD. As the
number of solitons decreases by steps, so does the oscillation amplitude of $%
\langle m_{y}\rangle (t)$, while the regime remains periodic on the long
timescale, covering the entire cycle of the metamorphoses. The corresponding
PSD exhibits pronounced peaks at the subharmonic response frequency $\omega
=\Omega /2$, as well as its odd multiples, similar to the regimes based on
single and double standard solitons, which are outlined above.

The boundaries between the existence regions of localized states plotted in
Fig. \ref{fig_ER_fftnvt} are well defined in most cases, except for the
upper-left corner, where the boundaries are fuzzy. The reason for the
fuzziness is that, as mentioned above, stability regions of the respective
solutions overlap, and the multistability takes place. As the attraction
basins are different for different areas in the $(\nu ,h_{0})$ plane, the
initial condition given by Eqs. (\ref{eq_init_cond_x}) evolves towards
different attractors. The fuzzy boundaries involve breathing solutions only,
suggesting that the true attraction basins may have a fractal structure, and
the attractors are chaotic \cite{Nieto_et_al_NL_2020,
Sanjuan_et_al_PRE_2020, Daza__Wagemakers_Sanjuan_EurophysLett_2023}.

In the rest of the region under the Arnold tongue, the system responds with
either a spatially uniform state or delocalized patterns.

\unskip

\subsection{The power spectral density}

\label{subsection_PSD}

Panel (b) in Fig. \ref{fig_ER_fftnvt} shows the number of peaks in the power
spectral density from $\omega =0$ to $\omega =3\Omega $ in the existence
regions of localized states in the $(\nu ,h_{0})$ plane. There is a large
region with a reduced number of peaks $(\leq 5)$. This region coincides with
the union of those where the system exhibits either a single soliton or
double standard ones in Fig. \ref{fig_ER_fftnvt}(a). The chart is in
agreement with the examples displayed in Figs. \ref{fig_Im_fft}(a,c).
Single- and double-soliton regimes with no breathing feature only three
peaks at $\omega =\tfrac{1}{2}\Omega $, $\omega =\tfrac{3}{2}\Omega $ and $%
\omega =\tfrac{5}{2}\Omega $ (with some additional tiny peaks arising as the
level of the numerical errors). On the other hand, the number of peaks is
suddenly increased in the breather-soliton regions. Most of the
single-breather-soliton region features the number of peaks $\sim 50-100$.
In most of the lower-left double-breather region, the results are similar.
In the upper-right double-breather region, four well-delimited zones can be
recognized, where the numbers of peaks are on the order of $1$, $10$ and $100
$. The last of these zones surround the region of fast-drifting localized
modes, where, the number of peaks is counted in hundreds and thousands, with
a maximum of $\approx 3200$ peaks. This is also consistent with the examples
displayed in Figs. \ref{fig_Im_fft} (b,d).

\subsection{The center-of-mass (CM) drift of solitons}

\label{subsection_cm}

\begin{figure}[tbp]
\includegraphics[width=\linewidth]{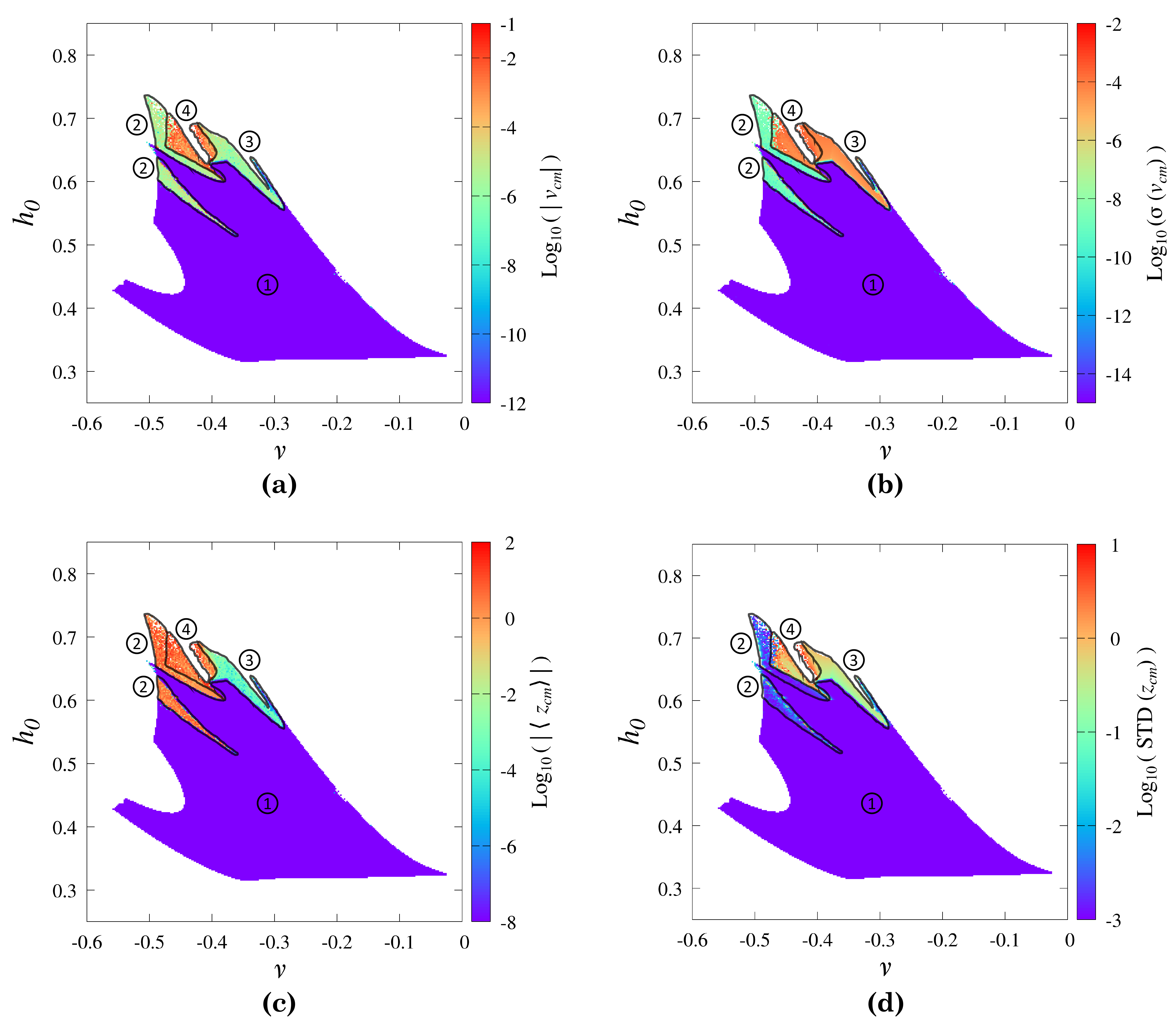}
\caption{(color online) Dynamical indicators related to the center of mass
(CM) of the soliton patterns, in the $(\protect\nu ,h_{0})$ plane, plotted
on the logarithmic scale. (\textbf{a}) The absolute value of the average CM
velocity. (\textbf{b}) The deviation of the CM velocity from its average
value. (\textbf{c}) The absolute value of the average CM position. (\textbf{d%
}) The standard deviation (STD) of the CM position. Four well-delimited
zones, designated by different colors, are identified, in terms of these
dynamical indicators.}
\label{fig_cm}
\end{figure}
\unskip

Figure \ref{fig_cm} summarizes indicators related to the CM mobility of the
soliton structure, at different values of $(\nu ,h_{0})$. For each point of
the map, the time series $z_{\mathrm{CM}}(t_{i})$, accounting for the CM
position, is extracted from the envelope data in the course of the
steady-state stage (i.e., at $t_{i}\in \lbrack 5000,6000]$), and the
following dynamical indicators are computed. The absolute value of the
average CM velocity, which is estimated as the slope of the linear fit for $%
z_{\mathrm{CM}}(t_{i})$, is shown in panel (a). The departure of the
velocity from its average is estimated as the standard deviation of the
linear-fit slope, considering $z_{\mathrm{CM}}(t_{i})$ terms as
normally-distributed random variables \cite{NumericalRecipes}. The deviation
is shown in panel (b). The absolute value of the time-average CM position
and its standard deviation are shown in Figs. \ref{fig_cm}(c) and (d),
respectively.

In Fig. \ref{fig_cm}, four well-delimited zones can be identified in terms
of these dynamical indicators. Zone $1$ (in purple) is the largest one,
encompassing all usual-single-soliton regimes, almost all
single-breathing-soliton ones, almost all regimes of usual double solitons,
and most cases of double breathing solitons. Here, the average soliton speed
is $\lesssim 10^{-12}$, the other indicators exhibiting extremely low values
too, which may be regarded as zero for all practical purposes. Zone $2$ is a
disconnected set at the upper-left corner of Fig. \ref{fig_cm}. It includes
a narrow edge occupied by single breathing solitons under (but disconnected
from) a narrow edge occupied by the usual double solitons, whose left end is
connected to the lower end of the edge occupied by single breathing
solitons. The solitons in this zone exhibit average speeds $\sim 10^{-6}$
with deviations $\sim 10^{-9}$. Such speeds imply a displacement $\sim
10^{-3}$ during the steady stage, which is much smaller than the mesh size $%
dz=1/6$ of the numerical scheme. Therefore, these solitons are also regarded
as non-drifting ones. However, the average displacement of the solitons from
$z=0$ ranges between $0.1$ and $10$. This occurs because, although the
solitons do not drift during the steady-state stage, they do so during the
preceding transient stage, thus producing an initial position at the
beginning of the steady-state stage far from $z=0$.

Zone $3$ is the right diagonal end of the breathing double-soliton region.
It has the same features as zone $2$ in terms of the average drift speed,
but the average CM position is closer to zero (with a distance $\sim
10^{-4}$), which means that the CM remains close to $z=0$ in the course of
both the transient and steady-state stages. On the other hand, both the CM
position and CM velocity exhibit larger, yet still small deviations, $\sim
10^{-1}$ and $10^{-5}$, respectively. These tiny fluctuations are caused by
the complexity of the breathing double-soliton structure. Zone $4$ is
located at the upper-left corner in Fig. \ref{fig_cm}, around the region of
the fast-moving localized modes, see Fig. \ref{fig_ER_fftnvt}(a). It is
occupied exclusively by the breathing double solitons, being the \emph{only
zone} that exhibits non-negligible drift. The respective drift speeds
exhibit average values in the range of $10^{-4}$ to $10^{-1}$, with a
maximum average speed found to be $0.032$, which allows the double breather
to travel the distance of $32.0$ during the steady-state stage. Considering
that the average CM positions have absolute values in the range of $1$ to $10
$, with a maximum of $39.28$, which implies that the double breathers remain
far enough from the wire's edges at $z=\pm 125$. All in all, some double
breathers exhibit the persistent drift with an almost constant velocity,
while others move with a sign-changing (oscillating) velocity.

\subsection{The micromagnetic energy}

\label{subsection_Q}

\begin{figure}[tbp]
\centering
\begin{subfigure}{0.485\textwidth}
        \includegraphics[width=\linewidth]{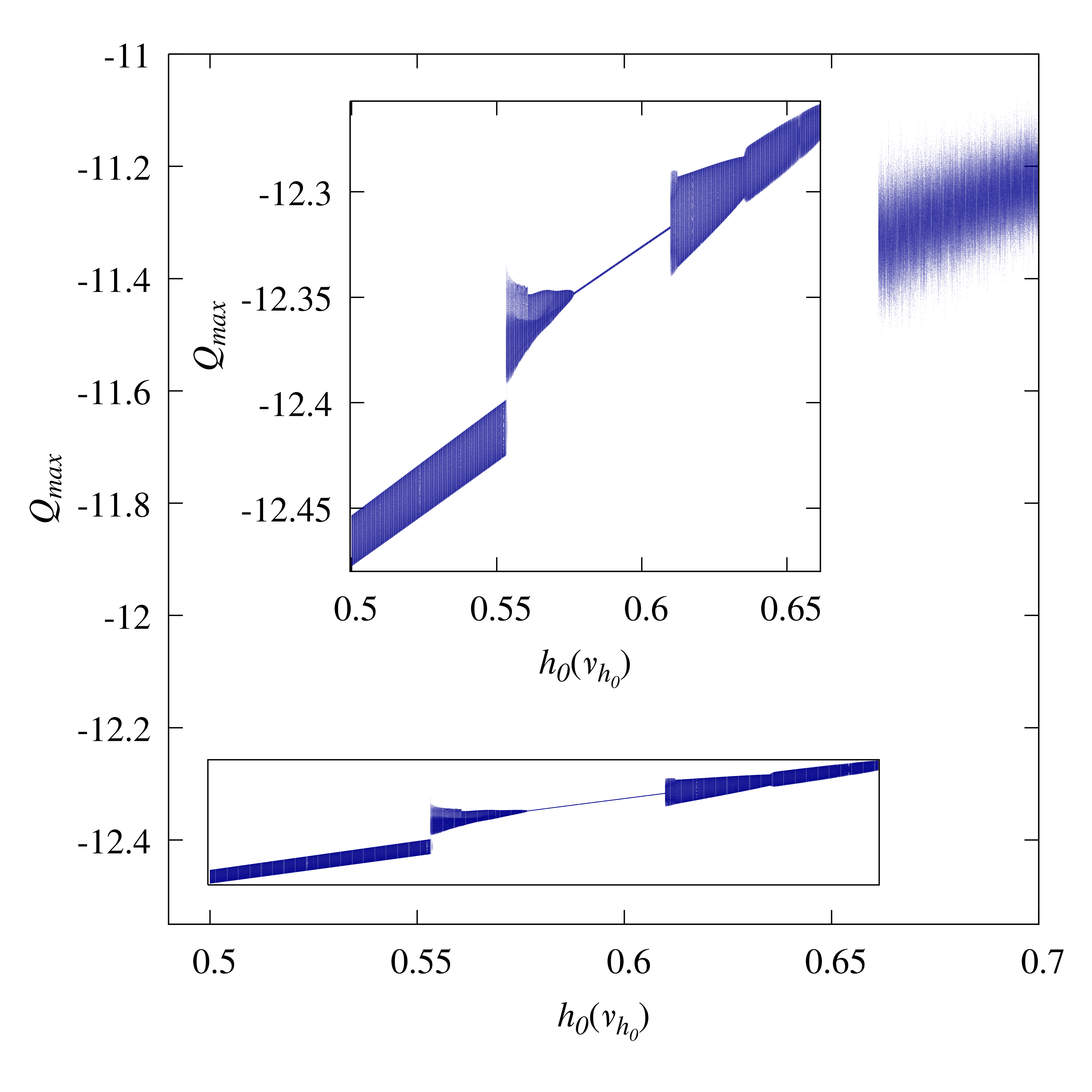}
        \caption{\centering Bifurcation diagram}\label{fig:qmax}
    \end{subfigure}
\begin{subfigure}{0.485\textwidth}
        \includegraphics[width=\linewidth]{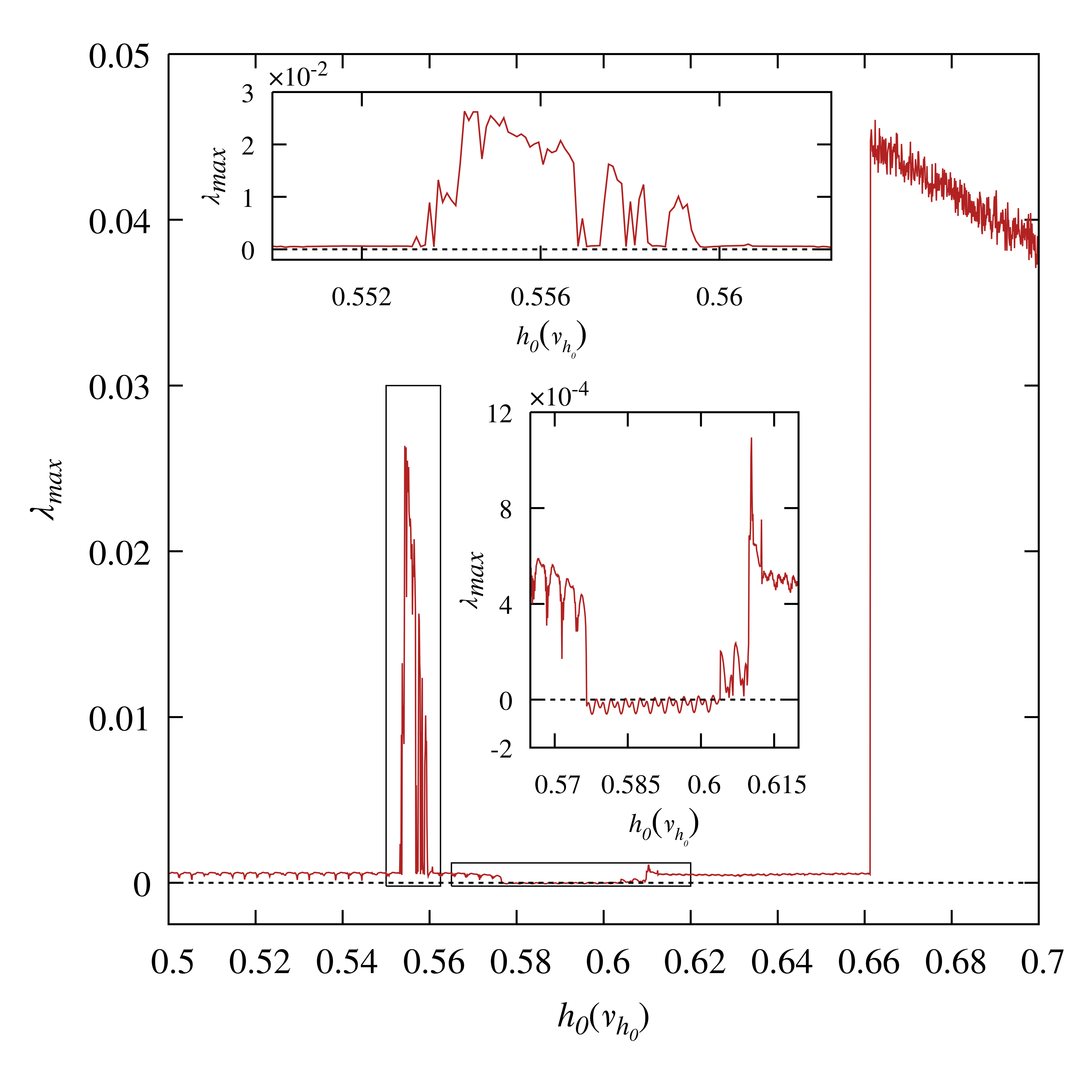}
        \caption{\centering Average value of the LLE}\label{fig:lya}
    \end{subfigure}
\caption{Bifurcation diagrams plotted along the line $h_{0}(\protect\nu %
_{h_{0}})=2.857\protect\nu _{h_{0}}+1.729$. (\textbf{a}) Local maximum
values of the micromagnetic free energy. (\textbf{b}) The average value of
the largest Lyapunov exponent. Iinsets in each panel zoom into the framed areas
of the corresponding diagrams.}
\label{fig_lyaqmax}
\end{figure}
\unskip

Figure \ref{fig_lyaqmax}(a) summarizes the computation of the maximum value
of the micromagnetic free energy along the line which was shown above in
Fig. \ref{fig_ER_fftnvt}, defined as $h_{0}(\nu _{h_{0}})=2.857\nu
_{h_{0}}+1.729$ and ranging between points $(\nu ,h_{0})=(-0.43,+0.5)$ and $%
(\nu ,h_{0})=(-0.36,+0.7)$. For $2048$ points uniformly distributed along
the line, the peak values of the time series for the energy were found.
Along the line, a number of segments can be identified in terms of the peak
energies. First, in the segment $0.5<h_{0}<0.553$, the peak energies
increase with a uniform slope of $\approx 0.98$, always lying in a band of
width $\Delta Q=0.025$. This segment is the one where the line passes
through the lower region of the breathing single solitons. At $h_{0}\approx
0.553$, there is a sudden increase in the peak energies, as the line is now
passing through the region of the breathing double solitons. The energy
increases here, as a longer section of the wire is now occupied by the
excited state. As $h_{0}$ grows, so do the peak energies, but the width of
the band decreases. Close to $h_{0}=0.579$, the peak-energy branch becomes a
line, which extends up to $h_{0}=0.609$ with a slope of $\approx 0.97$. This
makes sense because the respective segment passes through the region of the
standard (usual) double solitons. As the long-timescale dynamics remains
steady, the peak energies are identical to one another. At $h_{0}=0.609$,
the line enters the upper region populated by double breathing solitons, and
the peak energies again appear as bands with a width that tends to increase
with $h_{0}$. All in all, the peak energies keep increasing with $h_{0}$ at
a fairly constant rate. At $h_{0}=0.662$, the peak energies increase
suddenly from about $-12.3$ to $-11.3$, as the line enters the region where
the system switches into the subharmonic pattern, the entire wire being now
occupied by the excited state.

\subsection{The largest Lyapunov exponent (LLE)}

\label{subsection_LLE}

The LLE was computed for the same set of $2048$ points along the same line
plotted in Fig. \ref{fig_ER_fftnvt}). The results are summarized in Fig. \ref%
{fig_lyaqmax}(b).\ Similar to the peak energies, changes in the LLE along
the line can be associated with its segments traversing different existence
regions, see Fig. \ref{fig_ER_fftnvt}(a). For $h_{0}\in \lbrack 0.5,0.553]$,
in the region of the breathing single solitons, the LLE shows values $\sim
10^{-4}$. These small positive values imply that the system is weakly
chaotic in this case. Similar LLE values are observed for $h_{0}\in \lbrack
0.561,0.579]$, in the lower region of the breathing double solitons. Between
these two intervals, i.e., at $h_{0}\in \lbrack 0.553,0.561]$, the LLE
exhibits much larger values, up to $0.026$. This part of the line covers the
lower edge of the lower region of the breathing double solitons, where the
respective PSD exhibits significantly more peaks than in the rest of the
region (about $10$ times more, as shown in Fig. \ref{fig_ER_fftnvt}(b)).
This difference is related to the drastic increase in the LLE values in this
segment. For $h_{0}\in \lbrack 0.579,0.604]$, in the region of usual double
solitons, the LLE takes negative values $\sim -10^{-5}$. This means that the
usual double-soliton states are non-chaotic. For $h_{0}\in \lbrack
0.604,0.662]$, in the upper region of the breathing double solitons, the LLE
takes positive values $\sim 5\times 10^{-4}$. At $h_{0}>0.662$, in the
region of the subharmonic patterns, the LLE drastically increases to about $%
0.045$, which is expected, given the chaotic shape of these patterns
occupying the entire wire. All in all, the LLE analysis shows that the
dynamics of the usual solitons is regular (non-chaotic), while it is chaotic
for the breathing solitons, and subharmonic patterns are chaotic too. This
is consistent with the results for single solitons reported in Ref. \cite%
{Urzagasti2013}.

\section{Conclusions}

\label{section_Conclusions}

The subject of this work is the systematic study of dynamical regimes in the
one-dimensional rectilinear magnetic wire modeled by the LLG
(Landau-Lifshitz-Gilbert) equation with the Neumann's boundary conditions,
driven by the spatially uniform magnetic field, including DC and AC
components, which is directed perpendicular to the wire. The AC parametric
drive maintains magnetic excitations at the subharmonic frequency. By means
of the systematic numerical analysis, which includes the computation of PSD
(power spectral density) with the help of the Fourier transform, areas in
the parameter plane of the drive's amplitude and frequency detuning from the
parametric resonance are identified, in which the driven LLG equation
supports different robust localized dynamical states. These include usual
single- and double-soliton modes, single and double breathers (which emit
small-amplitude dispersive waves), fast-moving modes, and intermittent
multi-soliton complexes, which exhibit periodic metamorphoses with the
change in the number of solitons. The results reveal the presence of
multistability in the system, which includes the coexistence of stable localized
states and delocalized ones that cover the entire wire. Basic dynamical
characteristics of these states are found, such as the average
magnetization, LLE (largest Lyapunov exponent), and velocity of the CM
(center of mass) drift, in the case of the double breathers with
spontaneously broken inner symmetry. The states with negative and positive
values of LLE feature, respectively, regular and chaotic evolution.

The dynamical states addressed in the present work do not exhaust the variety 
of complex modes that the LLG equation may generate. More sophisticated states,
such as multisoliton bound states, may be a subject for further 
investigation.

It may also be relevant to extend the work by considering the model based on the
LLG equation with periodic boundary conditions, corresponding to the
ring-shaped magnetic wire. Besides, it may be interesting to study
collisions between the fast-moving modes. In these cases, a comparison with
the simplified model based on the PDNLS (parametrically-driven damped
nonlinear Schr\"{o}dinger) equation can help to establish the universality
of the phenomena. A challenging possibility is to develop a two-dimensional
version of the setting. Indeed, it is known that many models predict the
existence of stable two-dimensional solitons maintained by the balance of
gain and loss (see, e.g., Ref. \cite{NonlinDyn2023}).

\section*{Acknowledgments}

The work of B.A.M. was supported, in part, by the Israel Science Foundation through grant No. 1695/22. PD and DL acknowledge partial financial support from FONDECYT 1231020. We thank Mr. Manuel J. Suazo from University of of Tarapac\'{a} for his valuable advice to implement the parallelization in our numerical algorithms. DL acknowledges the hospitality of {\it Universit\'e C\^ote d'Azur} (Nice, France) and {\it Universidad Nacional de San Agust\'in} (Arequipa, Per\'u) where part of this work was written.  

\section*{Abbreviations}

The following abbreviations are used in this manuscript:\\

\noindent
\begin{tabular}{@{}ll}
LLG & Landau-Lifshitz-Gilbert\\
PSD & Power Spectral Density\\
LLE & Largest Lyapunov Exponent\\
CM  & Center of Mass
\end{tabular}

\bibliographystyle{unsrtnat}
\bibliography{Bibliography}  %%% Uncomment this line and comment out the ``thebibliography'' section below to use the external .bib file (using bibtex) .




% 	\bibitem{keshet2016prediction}
% 	Keshet, Renato, Alina Maor, and George Kour.
% 	\newblock Prediction-Based, Prioritized Market-Share Insight Extraction.
% 	\newblock In {\em Advanced Data Mining and Applications (ADMA), 2016 12th International 
%                       Conference of}, pages 81--94,2016.

% \end{thebibliography}

\end{document}